\documentclass[trackchanges, twocolumn]{aastex701}

\usepackage{xcolor}
\usepackage{booktabs}
\usepackage{multirow}
\usepackage{graphicx}
\usepackage{amsmath}
\usepackage{mathtools}

\newcommand{\ml}{\mathcal{L}}
\newcommand{\haloflow}{\texttt{HaloFlow}}
\newcommand{\mh}{$M_\mathrm{h}$}

\newcommand{\Simba}{\textsc{Simba}}
\newcommand{\Eagle}{\textsc{Eagle}}
\newcommand{\TNG}{\textsc{TNG}}

\newcommand{\eg}{\textit{e.g.,}}

\begin{document}

\title{HaloFlow II: Robust Galaxy Halo Mass Inference with Domain Adaptation}

\author[0000-0003-3418-2482]{Nikhil Garuda}
\affiliation{Department of Astronomy, The University of Texas at Austin, 2515 Speedway Boulevard, Austin, TX 78712, USA}
\affiliation{Cosmic Frontier Center, The University of Texas at Austin, Austin, TX, USA}
\altaffiliation{galactic.ai}
\email[show]{garuda@utexas.edu}

\author[0000-0003-1197-0902]{ChangHoon Hahn}
\affiliation{Department of Astronomy, The University of Texas at Austin, 2515 Speedway Boulevard, Austin, TX 78712, USA}
\affiliation{Cosmic Frontier Center, The University of Texas at Austin, Austin, TX, USA}
\altaffiliation{galactic.ai}
\email{}

\author[0000-0003-4758-4501]{Connor Bottrell}
\affiliation{International Centre for Radio Astronomy Research, University of Western Australia, 35 Stirling Hwy, Crawley, WA 6009, Australia}
\email{}

\author[0000-0001-9299-5719]{Khee-Gan Lee}
\affiliation{Kavli IPMU (WPI), UTIAS, The University of Tokyo, Kashiwa, Chiba 277-8583, Japan}
\affiliation{Center for Data-Driven Discovery, Kavli IPMU (WPI), UTIAS, The University of Tokyo, Kashiwa, Chiba 277-8583, Japan}
\email{}

\begin{abstract}
Precise halo mass (\mh) measurements are crucial for cosmology and galaxy formation. 
\haloflow~\citep{hahn_haloflow_2024} introduced a simulation-based inference (SBI) framework that uses state-of-the-art simulated galaxy images to precisely infer \mh. However, for \haloflow~to be applied to observations, 
it must be generalizable even when the underlying galaxy formation physics differ from those in the simulations on which it was trained. %
Without this generalization,  %
\haloflow~produces biased and overconfident \mh~posteriors when applied to simulations with different physics.
We introduce \haloflow$^{\rm DA}$, an extension of \haloflow~that integrates domain adaptation (DA) with SBI to mitigate these cross-simulation shifts. %
 Using synthetic galaxy images forward-modeled from the IllustrisTNG, EAGLE, and SIMBA simulations, we test two DA methods: Domain-Adversarial Neural Networks (DANN) and Maximum Mean Discrepancy (MMD).
Incorporating DA significantly reduces bias and improves calibration, with MMD achieving the most stable performance, lowering the normalized residual metric, $\beta$, by an average of 31\% and up to 57\% when trained and tested on different simulations.
Overall, \haloflow$^{\rm DA}$ produces more robust, less biased with similar precision, $M_h$ constraints  than the standard approach using the stellar-to-halo mass relation.
\haloflow$^{\rm DA}$~enables consistent, simulation-trained inference models to generalize across domains, establishing a foundation for robust \mh~inference from real HSC-SSP observations.

\end{abstract}

\keywords{}

\section{Introduction}
Robustly inferring the host dark matter halo masses of galaxies plays a critical role in cosmology and galaxy evolution studies.
The most massive halos that host galaxy clusters are promising probes of the 
expansion history of the Universe and the growth rate of structure~\citep[\eg][]{detf2006}. 
Host halos also play a major role in the evolution of galaxies~\citep[\eg][]{wechsler2018},
with halo mass playing the most dominant role in the galaxy-halo connection~\citep[\eg][]{tinker_are_2011, behroozi2019_um}.
Halo masses are also crucial for studying cosmic baryon distribution using 
fast radio bursts~\citep[FRBs;][]{lee_constraining_2022, lee2023_frb}, which probe 
the amount and extent of circumgalactic medium (CGM). Interpreting these FRB-based CGM measurements relies on the baryon-dark matter scaling relations, which quantify how baryon content varies with dark matter halo mass \citep{ayromlou_feedback_2023, sullivan_predicting_2024}, enabling constraints on the total baryon fraction in halos.

Despite its importance, the halo mass of an individual galaxy is not directly observable. Traditional approaches, therefore, rely on indirect methods. Weak and strong gravitational lensing provide relatively direct constraints on projected mass profiles, but high–fidelity measurements require deep, high–resolution imaging and large background source densities \citep{treu_strong_2010, hoekstra_masses_2013, bartelmann_weak_2016, mandelbaum_weak_2018, shajib_strong_2024, natarajan_strong_2024}. These requirements make lensing observationally expensive, limit its applicability to low–mass halos \citep{li_weak_2025}, and can restrict analyses to stacked measurements over ensembles of galaxies rather than precise constraints for individual systems \citep{putter_using_2005, yang_weak_2006}. %

A complementary set of techniques infers halo masses statistically from galaxy properties. Abundance-matching techniques relate galaxies to dark matter halos by requiring consistency between observed luminosity or stellar-mass functions and the halo mass function \citep{kravtsov_dark_2004, tasitsiomi_modeling_2004, vale_linking_2004, hearin_sham_2013, moster_galactic_2013}. 
While these methods have been highly successful in recovering the average stellar-to-halo mass relation and its scatter \citep{zaritsky_photometric_2022}, they constrain halo mass applied to an aggregated ensemble of galaxies.

A distinct class of approaches estimates halo masses of galaxy groups or clusters by leveraging group dynamics, especially in large spectroscopic surveys \citep{eke_galaxy_2004, yang_galaxy_2007, yang_galaxy_2009, robotham_galaxy_2011, wojtakGalaxyClusterMass2018b}. In these methods, group–finding algorithms are used to identify associations of galaxies, after which dynamical masses are estimated using the measured velocity dispersion and projected radius of group members through application of the virial theorem or related mass estimators \citep{heisler_estimating_1985, beers_measures_1990, robotham_galaxy_2011, tinker_are_2011, tinker_self-calibrating_2022, 2026A&C....5401011L}.
These methods provide valuable, often direct, constraints on halo mass; %
however, %
they are not designed to deliver well–calibrated posterior distributions of halo mass for individual galaxies.%

The rapid growth of wide, deep imaging surveys and realistic cosmological simulations has opened a complementary path: using machine learning (ML) to learn the mapping between galaxy observables and halo properties directly \citep{ntampaka_machine_2015, ntampaka_dynamical_2016, calderon_prediction_2019, villanueva-domingo_inferring_2022, wu_how_2024, garuda_estimating_2024}. In particular, simulation–based inference \citep[SBI; see][for recent review]{cranmer_frontier_2020} denotes a broad class of likelihood–free methods that use forward models to generate mock observations. Among the various SBI approaches, we focus on those based on neural density estimation, in which flexible conditional density models are trained on simulated data. In the variant known as neural posterior estimation \citep[NPE; ][]{papamakarios_fast_2018, greenberg_automatic_2019}, these networks directly approximate the posterior distribution of the parameters given the observations, $p(\theta \mid \mathbf{X})$.

\haloflow~\citep{hahn_haloflow_2024} uses NPE with normalising flows \citep{tabak_density_2010, tabak_family_2013} to infer the joint posterior of stellar and halo mass from photometric, morphological, and structural features extracted from synthetic galaxy images. 
\cite{hahn_haloflow_2024} demonstrated that \haloflow~can recover accurate and well–calibrated posteriors when trained and tested on galaxies drawn from the same cosmological hydrodynamical simulation. 
Moreover, it demonstrated that we can derive more precise $M_h$ constraints by leveraging imaging features beyond photometry.

However, like most ML approaches, \haloflow~inherits a key limitation: it is intrinsically dataset–specific. %
In practice, the network learns the mapping between features $\mathbf{X}$ and parameters $\theta$ implied by the training simulation, and implicitly assumes that the data it is applied to are drawn from the same underlying generative process (in–domain). This assumption is violated as soon as we move beyond the idealized ``train and test on the same simulation'' case. When a model trained on one simulation is applied either to a different simulation or, ultimately, to real observations, the mapping between $\mathbf{X}$ and $\theta$ encoded by the inference pipeline no longer matches the true generative process (out-of-domain). 
Each simulation we consider has different physical models of galaxy formation. 
None is precisely accurate to the physics of the \textit{real} Universe. %
This kind of domain shift constitutes model misspecification \citep{cannon_investigating_2022, huang_learning_2023, montel_tests_2025}. As we show later in this work, such shifts can drive \haloflow~to produce biased and over–confident posteriors for \mh.

To mitigate these effects, we seek techniques that enable an NPE trained on one to make reliable inferences in another. %
A common strategy is to modify the internal features learned by the network so that they become less sensitive to the choice of domain, while still retaining the information needed for precise inference. %
Unsupervised domain adaptation \citep[DA;][]{kouw_introduction_2019, farahani_brief_2020} provides a natural framework for this problem. In DA, one has labeled data in one or more source domains and unlabeled data in a target domain. The goal is to learn some feature representation of the data in which the distributions of training and test representations are statistically aligned, while preserving the information needed for the primary task. A variety of DA strategies have been developed, including adversarial methods such as Domain–Adversarial Neural Networks \citep[DANN;][]{csurka_domain-adversarial_2017} and distance–based approaches that minimize measures like the maximum mean discrepancy \citep[MMD;][]{gretton_kernel_2008} between feature distributions. DA has already seen successful applications in astronomy and cosmology, for example, in galaxy morphology classification \citep{ciprijanovic_deepadversaries_2022, ciprijanovic_deepastrouda_2023}, supernova classification \citep{vilalta_general_2019}, cosmological parameter inference \citep{roncoli_domain_2024}, and strong–lensing detection \citep{parul_domain_2025} and parameter estimation \citep{swierc_domain_2023, agarwal_neural_2025}.

In this work we bring DA into the \haloflow~framework to improve the robustness of \mh~inference under %
domain shifts. We use forward–modeled galaxy images from state–of–the–art cosmological hydrodynamical simulations: IllustrisTNG \citep{nelson_illustristng_2019}, EAGLE \citep{schaye_eagle_2015}, and SIMBA \citep{dave_simba_2019}-- processed with radiative–transfer and survey–realism pipelines \citep{bottrell_illustristng_2023, eisert_ergo-ml_2024} to mimic Hyper Suprime–Cam Subaru Strategic Program \citep[HSC-SSP;][]{aihara_third_2022} observations. We treat each simulation as a distinct domain and explicitly quantify how \haloflow’s performance degrades when trained on one simulation and evaluated on another. We then introduce \haloflow$^{\rm DA}$, a domain–adapted extension of \haloflow~in which an intermediate DA network maps observed features $\mathbf{X}$ into compressed, domain–invariant representations $c \mathbf{X}$ that are subsequently used as input to the NPE. By construction, these representations are encouraged to discard simulation–specific features while retaining information relevant for \mh. Although our ultimate goal is to apply \haloflow$^{\rm DA}$ to real survey data, in this paper we use multiple simulations as controlled proxies for the unknown mismatch between any given simulation and the real Universe, and we demonstrate that \haloflow$^{\rm DA}$ substantially improves out-of-domain performance.

We begin in \S\ref{sec:data} with a brief description of the hydrodynamical simulations and the forward–modeled synthetic images. In \S\ref{sec:methods}, we present \haloflow, introduce the two DA techniques (DANN and MMD), and define \haloflow$^{\rm DA}$, which combines \haloflow~with domain adaptation. We then present the results of applying \haloflow$^{\rm DA}$ in \S\ref{sec:results}, discuss their implications in \S\ref{sec:discuss}, and conclude in \S\ref{sec:conclusion}.

\section{Data} \label{sec:data}
The main goal of this work is to enable robust SBI of $M_*$ and \mh~from galaxy images. Achieving this requires realistic synthetic observations that incorporate both galaxy formation physics and observational systematics. 
To this end, we use forward-modeled galaxy images from state-of-the-art cosmological hydrodynamical simulations.
Furthermore, we want to demonstrate that \haloflow$^{\rm DA}$ is robust to domain shift so that we can ultimately apply it to observations. 
Therefore, we use simulations that adopt different sets of subgrid prescriptions, galaxy formation physics, and  resolution. %

\subsection{Cosmological Hydrodynamical Simulations} \label{subsec:methods_sims}
We use four large-scale cosmological hydrodynamical simulations as the basis for generating forward-modeled galaxy images: TNG50 and TNG100 from the IllustrisTNG suite, EAGLE Ref-L100N1504, and SIMBA m100n1024. %

{\bf IllustrisTNG} utilizes the moving-mesh magnetohydrodynamics code AREPO \citep{springel_e_2010}. 
We use TNG50 \citep{nelson_first_2019, pillepich_first_2019} and TNG100 \citep{springel_first_2018, pillepich_first_2018, naiman_first_2018, nelson_first_2018, marinacci_first_2018}, which span (51.7 Mpc)$^3$ and (110.7 Mpc)$^3$ volumes respectively. 
TNG50 has high spatial and mass resolution (baryon mass resolution of $M_{\rm b}{\sim}8.5 \times 10^4~M_\odot$) capable of resolving detailed internal galaxy structure, while TNG100 samples a larger volume at a coarser resolution ($M_b{\sim}1.4 \times 10^6~M_\odot$). 
Both runs adopt the same subgrid galaxy-formation model, including two-mode AGN feedback (thermal and kinetic) and kinetic stellar winds, although some subgrid parameter values are adjusted between TNG50 and TNG100 to account for the different resolution.

To leverage the strengths of both volumes, we combine the TNG50 and TNG100 datasets by applying systematic corrections to the TNG50 galaxy properties. Specifically, stellar masses in TNG50 are adjusted by aligning the stellar–halo mass relation (SHMR) with that of TNG100 over well-sampled mass ranges, following the mass-modulated rescaling procedure described in Appendix A1 of \citet{pillepich_first_2018}\footnote{This approach corrects for resolution-dependent discrepancies by applying a smooth, halo-mass-dependent correction, derived from the ratio of SHMRs between the two simulations, to the TNG50 stellar masses.}. Magnitudes and surface brightnesses are similarly corrected by applying the same magnitude corrections to Sersic magnitudes and related parameters. These corrections are derived from boxcar-averaged offsets between TNG50 and TNG100 as a function of halo mass, enabling smooth interpolation. This calibration step ensures that the combined \TNG~dataset presents a physically consistent 
distribution of galaxy properties.

{\bf EAGLE} \citep{schaye_eagle_2015, crain_eagle_2015} uses a modified smoothed particle hydrodynamics code, GADGET-3, with the ``Anarchy'' hydrodynamics scheme \citep{schallerEAGLESimulationsGalaxy2015}. We use the Ref-L100N1504 run with a (100 Mpc)$^3$ volume and $1504^3$ particles per species ($M_b{\sim}1.8 \times 10^6~M_\odot$). \Eagle~employs stochastic thermal stellar and AGN feedback, with subgrid parameters calibrated primarily to reproduce the $z \approx 0.1$ galaxy stellar mass function, galaxy sizes, and the black hole mass--stellar mass relation \citep{crain_eagle_2015}.

{\bf SIMBA} is built on the GIZMO \citep{hopkins_new_2015, hopkins_new_2017} code in meshless finite mass mode \citep{dave_simba_2019}. We uses the flagship m100n1024 run, a (100 $h^{-1}$ Mpc)$^3$ volume with $1024^3$ particles per species ($M_b{\sim}1.8 \times 10^7~M_\odot$). \Simba~incorporates an H$_2$-based star formation model, kinetic stellar winds, and a multi-mode AGN feedback scheme with high-velocity jets and X-ray heating, leading to distinct gas and halo properties compared to \TNG~and~\Eagle.

In this work, we focus exclusively on central galaxies at redshift $z=0.1$ and $M_* > 10^{10} M_\odot$, selected from each simulation using the Friends-of-Friends (FoF) group catalogs. Central galaxies are identified as the primary subhalo %
within each FoF group, determined using the SUBFIND algorithm \citep{springel_populating_2001}. For each central galaxy, we extract its $M_*$ and host \mh~from the simulation outputs. %

\subsection{Realistic Synthetic Images} \label{subsec:forward_model}
We use publicly released synthetic galaxy images\footnote{https://www.tng-project.org/explore/gallery/bottrell23/} from the simulations described above. These images are produced with the forward-model described in \cite{bottrell_illustristng_2023}, adapted to work across multiple cosmological simulation suites \citep{eisert_ergo-ml_2024}. This forward-model yields realistic, observation-like data for use in SBI. The image-generation workflow comprises two stages: (1) dust-attenuated, noise-free rendering with \texttt{SKIRT}, and (2) observational realism that injects simulated galaxy light into HSC cutouts.

\begin{figure*}
    \centering
    \includegraphics[width=0.24\linewidth]{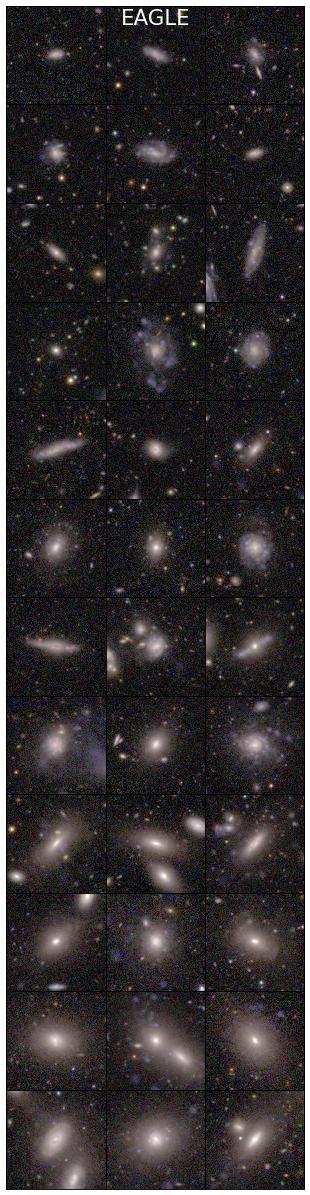}
    \includegraphics[width=0.24\linewidth]{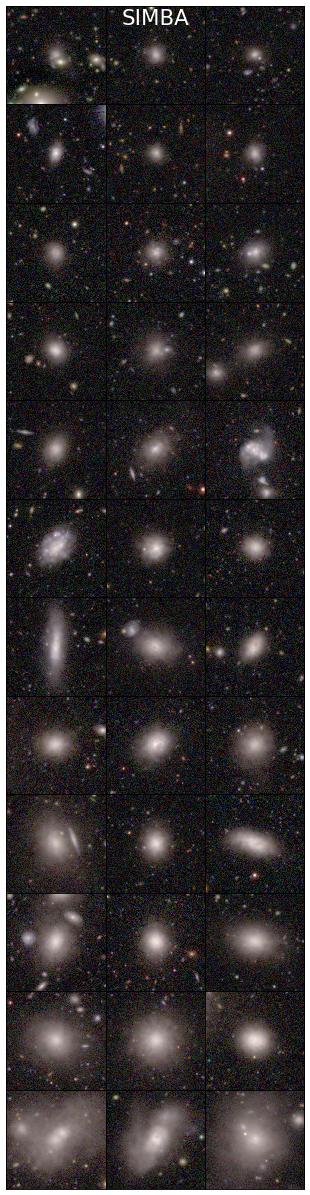}
    \includegraphics[width=0.24\linewidth]{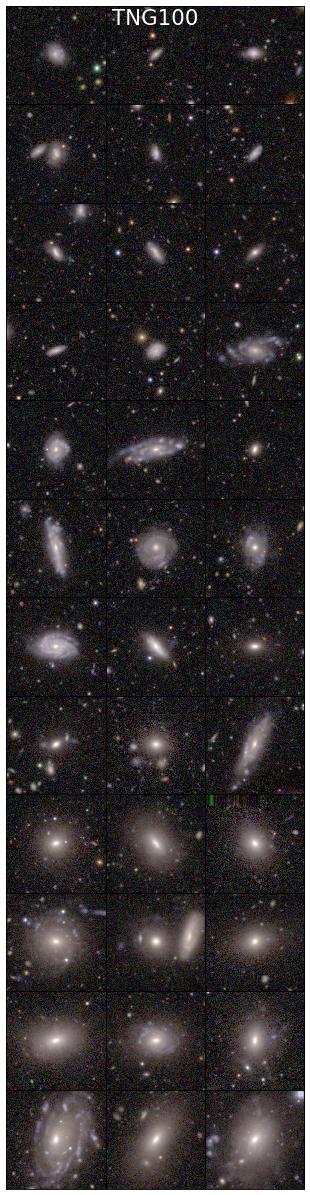}
    \includegraphics[width=0.24\linewidth]{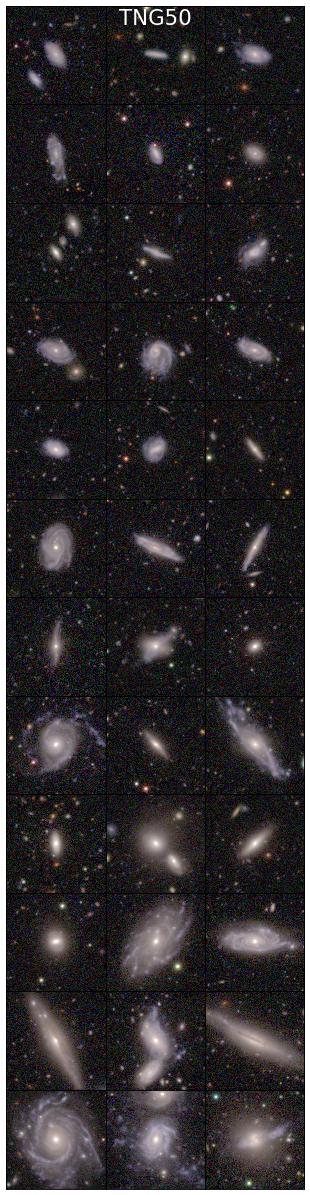}
    \caption{\textbf{HSC mock images of simulated galaxies} from the z $\approx$ 0.1 snapshot of \Eagle, \Simba, TNG100, and TNG50 (left to right) simulations. 
     36 central galaxies with stellar masses $10 < \log(M_*/M_\odot) < 11.6$ are shown. Stellar masses increase from upper to lower rows. Individual panels are 120 kpc (63 arcsec) across. RGB colours derive from the HSC $gri$ optical bands using arcsinh scaling.}
    \label{fig:simulated_gal_imgs}
\end{figure*}

In this framework, noise-free synthetic images are generated using the \texttt{SKIRT} dust radiative transfer code\footnote{https://skirt.ugent.be/} \citep{baes_efficient_2011, camps_skirt_2015, camps_skirt_2020}. For each central galaxy at $z = 0.1$, star and gas particle data are extracted within a spherical volume centered on the galaxy’s FoF halo, capturing its physical and environmental context. The radiative transfer simulation computes the propagation of starlight through a dusty medium, accounting for absorption, scattering, and thermal re-emission.

Idealized synthetic images in the HSC \textit{grizy} bands \citep{kawanomoto_hyper_2018} are produced, along with several auxiliary filters spanning 0.3–5 $\mu$m. To account for orientation-dependent effects and enhance the diversity of the dataset, each galaxy is rendered along four lines of sight, arranged in a tetrahedral configuration.

Stellar emission is modeled using the \citet{bruzual_stellar_2003} stellar population synthesis templates with a \citet{chabrier_galactic_2003} initial mass function (IMF) for stars older than 10 Myr. For younger stellar populations ($< 10$ Myr), the MAPPINGS III SED library \citep{groves_modeling_2008} is adopted, which includes nebular emission, dust continuum, and line emission from H$_{\rm II}$ regions and photodissociation regions.

Dust is modeled using a metallicity-scaled dust-to-gas prescription based on \citet{remy-ruyer_gas--dust_2014} and \citet{popping_dust-continuum_2022}, assuming a Milky Way dust grain size distribution and composition. While only some simulations (\eg~SIMBA) explicitly track dust particles, this ensures a uniform and realistic attenuation treatment across all simulations.

To emulate observational conditions, the survey systematics of the HSC-SSP Public Data Release 3 \citep{aihara_third_2022} are applied to the noise-free synthetic images. Using an adapted version of \texttt{RealSim}\footnote{https://github.com/cbottrell/RealSim} \citep{bottrell_deep_2019}, simulated galaxy images are inserted into real HSC background cutouts, with realistic point spread functions, spatial rebinning, pixel-level blending, and flux calibration applied. The resulting HSC-like mock images statistically reflect the depth, seeing, and crowding of real galaxy survey data.

Figure~\ref{fig:simulated_gal_imgs} shows %
HSC mock images for randomly selected galaxies at $z=0.1$ from \Eagle, \Simba, TNG100, and TNG50 (left to right) with increasing stellar mass (top to bottom). 
These images illustrate the morphological diversity predicted by each simulation suite with consistent observational realism.

\begin{table}
    \caption{Observational features ($\mathbf{X}$) of galaxy images used to infer $M_*$ and $M_h$ in \haloflow$^{\rm DA}$}
    \label{tab:breakdown_features}
    \begin{tabular}{c|c}
    \toprule
    \textbf{Observations} & \textbf{Parameters} \\
    \midrule
    $X_{\text{mag}}$ & HSC \textit{grizy} magnitudes \\
    $X_{\text{morph}}$ & $R_{\text{eff}}$, $n$, $b/a$ \\
    $X_{\text{extra}}$ & $CAS$, $G$, $M_{20}$, $\mu_{\text{1 kpc}}$, $A_{\text{res}}$ \\
    \bottomrule
    \end{tabular}
\end{table}

Photometric and morphological features are extracted using \texttt{ProFuse} \citep{robotham_profuse_2022}. Source identification and segmentation are performed with \texttt{ProFound} \citep{robotham_profound_2018} on a \textit{grizy} stacked image, after which \texttt{ProFit} \citep{robotham_profit_2017} is used to fit a single-component Sérsic profile to the target galaxy independently in each band. We additionally measure a suite of non-parametric morphology statistics. These measurements form the observational data vector $\mathbf{X}$ used in SBI. The data vector $\mathbf{X}$ is the combination of $X_{\text{mag}} + X_{\text{morph}} + X_{\text{extra}}$.

HSC \textit{grizy} magnitudes come from \texttt{ProFound} photometry. $R_{\text{eff}}$ and $n$ and $b/a$ are the Sérsic effective radius, characteristic index, and semi minor-to-major axis ratio. $CAS$ parameters are concentration, asymmetry, and smoothness \citep{conselice_relationship_2003}. $G$ is the Gini coefficient and $M_{20}$ is the second-order moment of the brightest pixels containing 20 percent of the galaxy flux \citep{lotz_new_2004}. $\mu_{\text{1 kpc}}$ is the mean surface brightness within 1kpc of the galaxy centre, $A_{\text{res}}$ is residual asymmetry \citep{simard_catalog_2011}. 
The observation features are listed in Table \ref{tab:breakdown_features}.
We note that for our purposes of demonstrating \haloflow$^{\rm DA}$, we do not include the total measured luminosity, $L_{\text{sat}}$, and number, $N_{\text{sat}}$, of satellites brighter than $M_r < -18$ for each individual group, which were included in the previous \haloflow~work.

The final dataset with four line of sight per galaxy includes 2,136 galaxies from TNG50\ and 15,364 from TNG100\ (total 17,500 for \TNG), along with 8,888 galaxies from \Eagle\ and 8,831 from \Simba. 
The dataset is divided into training and testing sets using a 90-10 split. 

\section{Methods} \label{sec:methods}
In this section, we describe our methodology for inferring \mh~from galaxy observations using SBI. Specifically, we describe how we incorporate DA into the \haloflow~framework. %
SBI methods like \haloflow~can recover accurate \mh~posteriors in in-domain settings, where training and test data are drawn from the same simulation. In such cases, the NPE encodes a specific galaxy–formation model and set of physical and numerical assumptions that are consistent across train and test samples. 
This assumption often breaks down in out–of–domain scenarios, where there is a domain shift between the training and test samples (\eg \TNG~$\Rightarrow$~\Eagle).

To address this, we introduce DA as an intermediate step before performing SBI. DA techniques aim to reduce differences across training and test domains --- in this case, across simulations with different galaxy formation physics --- to improve generalisation and enable more robust inference. We first train a DA network to map the observational features $\mathbf{X}$ into compressed, domain-invariant representations $c \mathbf{X}$, and then use $c \mathbf{X}$ as input to the \haloflow~SBI framework to learn $p(M_*, M_h \mid c \mathbf{X})$.

In this work, we use the different simulations as training and test domains. That is, we treat different cosmological simulations (\TNG, \Eagle, \Simba) as distinct domains and use them as a controlled testbed to study the impact of DA. Our working hypothesis is that techniques which improve generalization across simulations will also improve generalization to observational data.
We refer to our domain-adaptive SBI framework as \haloflow$^{\rm DA}$. In the subsections that follow, we detail its components: the SBI framework (\S \ref{subsec:haloflow}) and our DA strategy (\S \ref{subsec:da}).

\subsection{\haloflow} \label{subsec:haloflow}
\haloflow~is a SBI framework used to infer the posterior $p(\theta \mid {\mathbf X})$ where $\theta = \{M_*, M_h\}$ represents stellar and halo mass, and ${\mathbf X}$ denotes a set of features extracted from synthetic galaxy images.
SBI\footnote{also known as implicit likelihood inference}, enables posterior inference without requiring an explicit likelihood function. Instead, it leverages forward simulations that generate synthetic data ${\mathbf X}$ conditioned on physical parameters $\theta$. These forward models implicitly define the likelihood by simulating the data generation process from physical parameters to observable features. 

\haloflow~is based on NPE using ``normalising flow'' models. Flows use neural networks that maps complex target distributions to a simple base distribution, $\pi(z)$ via a sequence of invertible and bijective transformations, $f:x\rightarrow z$. The target distribution here is the posteriors $p(\theta \mid {\mathbf X})$, and $f$ is designed to be invertible and have a tractable Jacobian. By doing this, we can evaluate the posterior from $\pi(z)$ by change of variables. We choose a multivariate Gaussian for $\pi(z)$ as it makes the posterior easy to sample and evaluate. \haloflow~uses masked autoregressive flows \citep[MAFs;][]{papamakarios_masked_2018}, a type of normalizing flow that model a complex distribution using a series of invertible autoregressive transformations.

We train a flow, $q_\phi$, with hyperparameters, $\phi$, to best approximate $q_\phi(\theta~|~\mathbf{X}) \approx p(\theta~|~\mathbf{X})$. For each galaxy $\{(\theta_i, \mathbf{X}_i) \}$ in the training set, we maximise the total log-likelihood $\Sigma_i \log q_\phi(\theta_i~|~\mathbf{X}_i)$. This is equivalent to minimizing the Kullback–Leibler divergence between $p(\theta~|~\mathbf{X})p(\mathbf{X})$ and $q_\phi(\theta~|~\mathbf{X})p(\mathbf{X})$.
To determine the final normalising flow, \haloflow~employs an ensemble of independently initialized and trained flows for the final NPE for improved  accuracy~\citep{lakshminarayanan_simple_2019, alsing_fast_2019}.

$q_\phi$ implicitly encodes a prior, $p(\theta)$, which is set by $M_*$ and \mh~distributions of the training data. Without correction, this implicit prior mirrors the underlying stellar and halo mass functions, biasing inferred posteriors toward low $M_*$ and low \mh, where galaxies are more abundant. We correct this prior and impose uniform priors in $\log M_*$ and $\log M_{\rm h}$ using the \citet{handley_maximum-entropy_2019} maximum entropy prior method. In practice, for each posterior sample, $\theta^\prime \sim p(\theta^\prime \mid \mathbf{X})$, we apply an importance weight, $1/p(\theta^\prime)$ to ensure that we have uniform priors on $M_*$ and \mh.

\citet{hahn_haloflow_2024} estimated $p(\theta')$ using a binned halo mass function and linearly interpolating it to evaluate at each $\theta'$. %
The halo mass function is sparsely populated at the high-mass end, leading to noisy estimates of $p(\theta^\prime)$ and consequently large, unstable importance weights, further exacerbated by increased flow uncertainty in the posterior tails. 
To stabilise the correction, we fit a single Schechter profile to the halo mass function and use this smooth analytical form to model $p(\theta^\prime)$. Additionally, before applying the prior reweighting, we discard posterior samples lying beyond the 0.5th and 99.5th percentiles of the NPE posterior. This removes poorly constrained extreme tail samples that would otherwise generate pathological weights. Together, these modifications yield numerically stable, prior-corrected posteriors (see Appendix~\ref{app:prior_correction} for details).

In this work, we adopt the \haloflow~framework and maintain the same parameter space as in \citet{hahn_haloflow_2024}. We also use the same network architecture and hyperparameter optimization procedure as \citet{hahn_haloflow_2024}. We extend the original methodology to incorporate a broader range of galaxy features extracted from synthetic images across multiple simulations (see \S \ref{subsec:methods_sims} and Table~\ref{tab:breakdown_features}).

\subsection{Domain Adaptation} \label{subsec:da}
SBI methods rely on the assumption that the training and test data are drawn from the same distribution. However, this assumption is often violated in practice, particularly when flows trained on synthetic data are applied to real observations. Even across different simulations, variations in galaxy-formation physics, numerical resolution, numerical scheme and calibration can lead to systematic differences in the observable properties of galaxies. If unaddressed, these discrepancies lead to model misspecification.

Since the true generative process of the Universe in unknown, some level of model misspecification is unavoidable. To mitigate its effects, we introduce DA as an intermediate step before performing SBI. The goal of DA is to map galaxy features ${\mathbf X}$ into a representation space where their distributions are statistically aligned across different domains, $c \mathbf{X}$ -- in our case, across different simulations. By learning domain-invariant features, the posterior estimator becomes less sensitive to simulation-specific artifacts, improving robustness and generalisation.

\begin{figure*}
    \centering
    \includegraphics[width=0.7\linewidth]{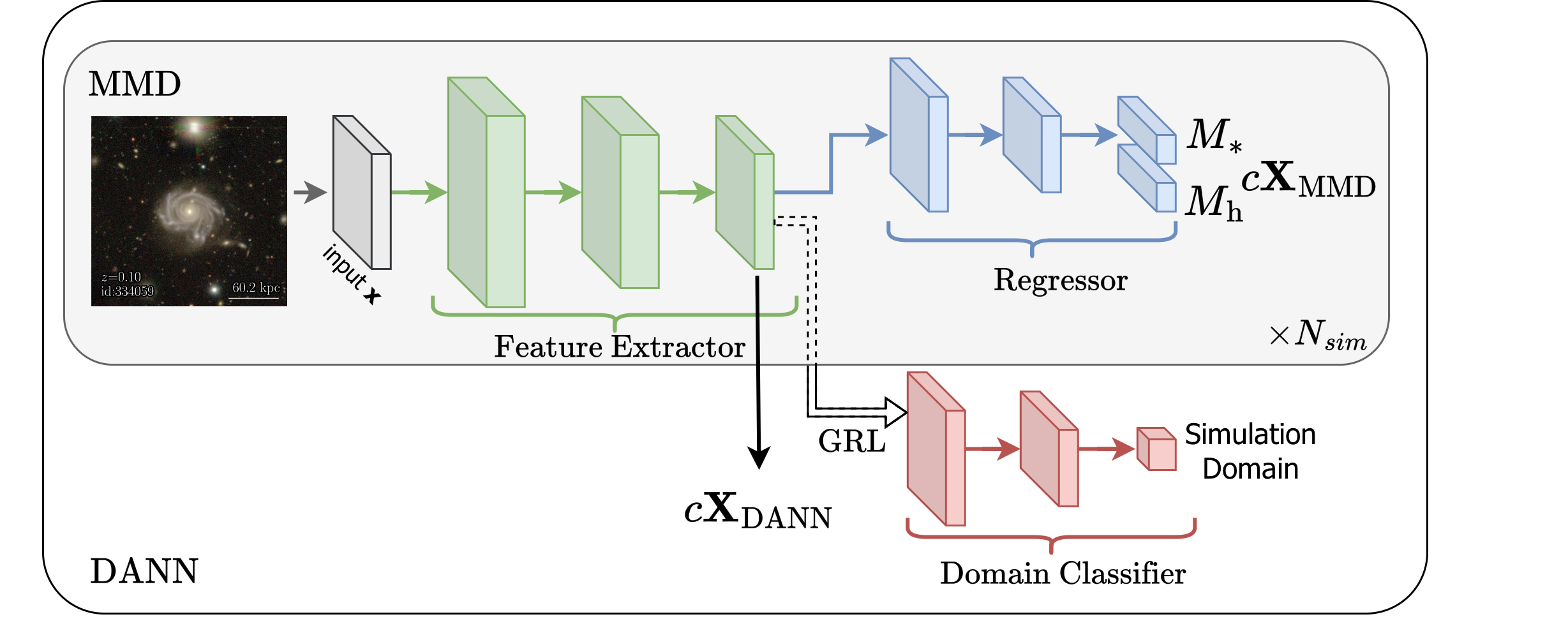}  
    \caption{Schematic of the DA frameworks used in this work. Images of galaxies are preprocessed into input features, $\mathbf{X}$, which are passed through a feature extractor (green), and a regressor (blue) to predict $M*$ and \mh~for each simulated galaxy. The dark gray panel (MMD) applies a MMD loss in the latent feature space, while the white panel (DANN) routes the extracted features through a gradient reversal layer to a domain-classifier network (red) that predicts which simulation the galaxy from and encourages domain-invariant features. The compressed features, $c \mathbf{X}$, used in \haloflow$^{\rm DA}$ are taken from the feature-extractor (32-dimensional) for the DANN runs, and from the label-predictor output (2-dimensional, $(M_*, M_{\rm h})$) for the MMD runs.}
    \label{fig:da_diagram}
\end{figure*}

In this work, we study two unsupervised DA techniques: (1) Domain Adversarial Neural Networks \cite[DANN, \S \ref{sec:dann}][]{csurka_domain-adversarial_2017}, which use adversarial learning to promote domain invariance, and Maximum Mean Discrepancy \cite[MMD, \S \ref{sec:mmd}][]{gretton_kernel_2008}, which directly minimizes distributional differences between domains using kernel-based distance metrics.
In Figure~\ref{fig:da_diagram}, we show a schematic overview of our DA setup. Together, these components define the DA extension of \haloflow$^{\rm DA}$. 
The original \haloflow~pipeline is applied not to the raw features $\mathbf{X}$ but to compressed, DA-informed representations $c\mathbf{X}$ produced by the DANN or MMD networks. 

\subsubsection{Domain Adversarial Neural Networks} \label{sec:dann}
DANNs are a technique for unsupervised DA that aims to address model misspecification through adversarial training. The goal is to learn a feature representation that is \textit{invariant} across different simulations, while still retaining the information needed to predict physical parameters, $\theta$.
DANN introduces an adversarial training setup composed of three parts: a ``feature extractor'' that maps ${\mathbf X}$ into a lower-dimensional latent representation $c \mathbf{X}$; a ``regressor'' that uses the latent representation to estimate $\theta$; and a ``domain classifier'' that tries to determine which simulation ${c \mathbf X}$ came from.
Crucially, the feature extractor is optimised adversarially: it is updated so as to make the domain classifier's task harder. This forces the extractor to learn domain-invariant features that contain physical information relevant for inference but suppress simulation-specific information.

To implement this adversarial setup, we use a gradient reversal layer (GRL). This layer behaves as the identity function during the forward pass, but multiplies the gradient by a negative scalar during backpropagation. As a result, the feature extractor is updated in a direction that maximally confuses the domain classifier, i.e., it attempts to remove the domain-specific information in the learned features.

The overall loss function is:
\begin{equation} \label{eq:loss}
    \ml = \ml_{\text{reg}} + \lambda ~\ml_{\text{DA}}
\end{equation}
where $\ml_{\text{reg}}$ is the regression loss; $\ml_{\text{DA}}$ is the domain classification loss; and $\lambda$ controls the relative strength of the domain-adversarial signal during training.

We define the regression loss as a weighted mean squared error (MSE):
\begin{equation} \label{eq:reg_loss}
\ml_{\text{reg}} = \sum_{i=1}^N w_i\, ||\hat\theta_i - \theta_i||^2
\end{equation}
where $\hat \theta_i$ is the predicted parameter vector, $\theta_i$ is the ground truth from the simulation. The weighting is introduced to prevent the model from being dominated by regions of parameter space that are overrepresented in the training data, particularly low-mass halos and to ensure adequate performance across domains across the full mass range. Without the weighting, the network would preferentially optimise accuracy where data are most abundant, i.e., at the expense of high-mass. %
The per-sample weight, $w_i$, therefore up-weights underrepresented domains and halo-mass ranges while down-weighting overrepresented ones; $w_i$ is the product of a domain-balancing term and a mass-dependent term based on an inverse Schechter-like function of halo mass (see Appendix~\ref{app:weights} for details).

The domain classification loss is a categorical cross-entropy: 
\begin{equation}
\ml_{\text{DA}} = - \sum_{i=1}^N \sum_{j=1}^K y_{ij} \log(p_{ij})
\end{equation}
where $K$ is the number of simulations, $y_{ij}$ is a one-hot encoded vector indicating the true simulation source, $p_{ij}$ is the predicted probability that sample $i$ came from simulation $j$. This loss is minimised by the domain classifier but maximised by the feature extractor via GRL. This adversarial objective ensures that the learned features do not encode which simulation a galaxy came from.

To stabilize training, we gradually increase the strength of the adversarial signal during training. Specifically, we set:
\begin{equation} \label{eq:lambda_dann}
\lambda(p) = \frac{2}{1 + \exp(-4.5 \cdot p)} - 1, \quad \text{where } p = \frac{\text{epoch}}{\text{total}} 
\end{equation}
This function smoothly increases $\lambda$ from 0 to 1 over the course of training, allowing DANN to first focus on learning the regression task before being strongly penalized for domain information. This schedule was found empirically to stabilize training and improve final inference performance (see \S\ref{sec:discuss} for further details).

For the compressed representation of observed features $c\mathbf{X}$ used as input for our SBI, we use the 32-dimensional $z$ output of the DANN feature extractor. %

\subsubsection{Maximum Mean Discrepancy (MMD)} \label{sec:mmd}
MMD is a kernel-based method for unsupervised DA. It provides a non-adversarial approach to learning domain-invariant features by directly minimizing a statistical distance between feature distributions from different simulations.

In our setup, MMD is used to reduce differences in the \textit{latent} features extracted by the network across simulations with different galaxy formation models.
An MLP feature extractor maps the observational feature vector $\mathbf{X}$ to a 32-dimensional representation $\mathbf{z}$ (see Appendix~\ref{app:arch}). A separate regressor then maps $\mathbf{z}$ to the two outputs $(M_*, M_{\rm h})$.
By minimizing this discrepancy between the $\mathbf{z}$ distributions of different simulations, the model is encouraged to focus on physical properties shared across simulations, rather than features tied to simulation-specific details, which can improve generalization to out-of-domain data. 

The core idea behind MMD is to embed the $\mathbf{z}$ distributions from different simulations into a reproducing kernel Hilbert space (RKHS), where comparisons between distributions can be made using kernel functions. In this space, MMD measures the squared distance between the mean embeddings of two distributions; if the MMD is small (close to zero), the corresponding feature distributions are considered statistically aligned.

We adopt a Gaussian Radial Basis Function (RBF) kernel to measure similarity between pairs of latent features,
\begin{equation}
k(\mathbf{z}, \mathbf{z}') = \exp\left( - \frac{\lVert \mathbf{z} - \mathbf{z}' \rVert^2}{2\sigma^2} \right),
\end{equation}
and fix the kernel bandwidth to $\sigma = 1.0$ and the MMD weight to $\lambda = 0.5$ for all experiments. These values were chosen based on a small grid search over $(\lambda, \sigma)$, which showed that the validation losses are relatively insensitive to modest variations in either parameter (Appendix~\ref{app:train}).

During training, we consider three simulations simulataneously: two source domain, $A$ and $B$, and one target domain, $C$. We compute latent features, $\mathbf{z}_A, \ \mathbf{z}_B, \ \mathbf{z}_C$ by passing $\mathbf{X}_A$, $\mathbf{X}_B$, and $\mathbf{X}_C$ through a shared feature extractor. The DA loss is then defined as the sum of the squared MMD distances between the target and each source,
\begin{equation}\label{eq:loss_mmd}
    \mathcal{L}_{\text{DA}} = \mathrm{MMD}^2(\mathbf{z}_A, \mathbf{z}_C) + \mathrm{MMD}^2(\mathbf{z}_B, \mathbf{z}_C),
\end{equation}
where each $\mathrm{MMD}^2$ term is computed using the RBF kernel above. This loss encourages the latent features of both training simulations to align with those of the target simulation. The three-way setup reduces the risk that the network overfits to a single simulation and provides a simple proxy for the situation in which real observations occupy an intermediate regime between simulations.

The final training objective for the MMD model has the same form as Eq.~\ref{eq:loss}:
\begin{equation}
    \mathcal{L} = \mathcal{L}_{\text{reg}} + \lambda\,\mathcal{L}_{\text{DA}}.
\end{equation}
The regression loss $\mathcal{L}_{\text{reg}}$ is defined as in Eq.~\ref{eq:reg_loss}, using only the two training simulations ($A$ and $B$) and the same per-sample weights $w_i$ described in Appendix~\ref{app:weights}. 
We emphasize that the test simulation $C$ enters only through the MMD term; hence, we do not use any $M_*$ or $M_h$ information of the test simulation. 
We fix $\lambda = 0.5$, which we find provides a good balance between fitting the regression task and enforcing cross-domain alignment. 

For $c\mathbf{X}$ used as input to our SBI, we use the 2D regressor output. 
Implementation details of the MMD architecture and training hyperparameters are described in Appendix~\ref{app:arch}.

\subsubsection{\haloflow$^{\rm DA}$: Domain-Adapted Posterior Inference}

For test simulation $C$, we designate the two other simulations, $A$ and $B$, as training domains. A DA model (DANN or MMD) is then trained using labeled samples from $A$ and $B$ and unlabeled samples from $C$. 
Once trained, the DA network is frozen and used purely as a feature extractor. 
We apply the frozen DA network to galaxies from simulations $A$ and $B$ to obtain compressed, domain-adapted representations $c\mathbf{X}_A$ and $c\mathbf{X}_B$, and we likewise compute $c\mathbf{X}_C$ for the target simulation. For the DANN runs, $c\mathbf{X}$ is taken to be the 32-dimensional latent output of the feature extractor; for the MMD runs, $c\mathbf{X}$ is taken to be the 2-dimensional regressor output. %

We then train \haloflow~in two separate NPEs: one using training pairs $(c\mathbf{X}_A,\theta_A)$, and the other using $(c\mathbf{X}_B,\theta_B)$. The key difference relative to the standard setup is that, instead of conditioning on the raw features $\mathbf{X}$ listed in Table~\ref{tab:breakdown_features}, \haloflow$^{\rm DA}$ learns posteriors conditioned on the compressed representations $c\mathbf{X}$.
Finally, we evaluate each trained NPE on the held-out test simulation $C$. 
$C$ is never used for training the NPE. It enters only through the DA stage as an unlabeled target domain. 

This setup provides a controlled testbed for assessing whether DA can improve generalization to out-of-domain data. In particular, we can examine how DA impacts the calibration, accuracy, and uncertainty of the resulting posteriors across different choices of train-test simulation combinations $(A, B \rightarrow C)$.

\begin{figure*}
    \centering
    \includegraphics[width=0.8\linewidth]{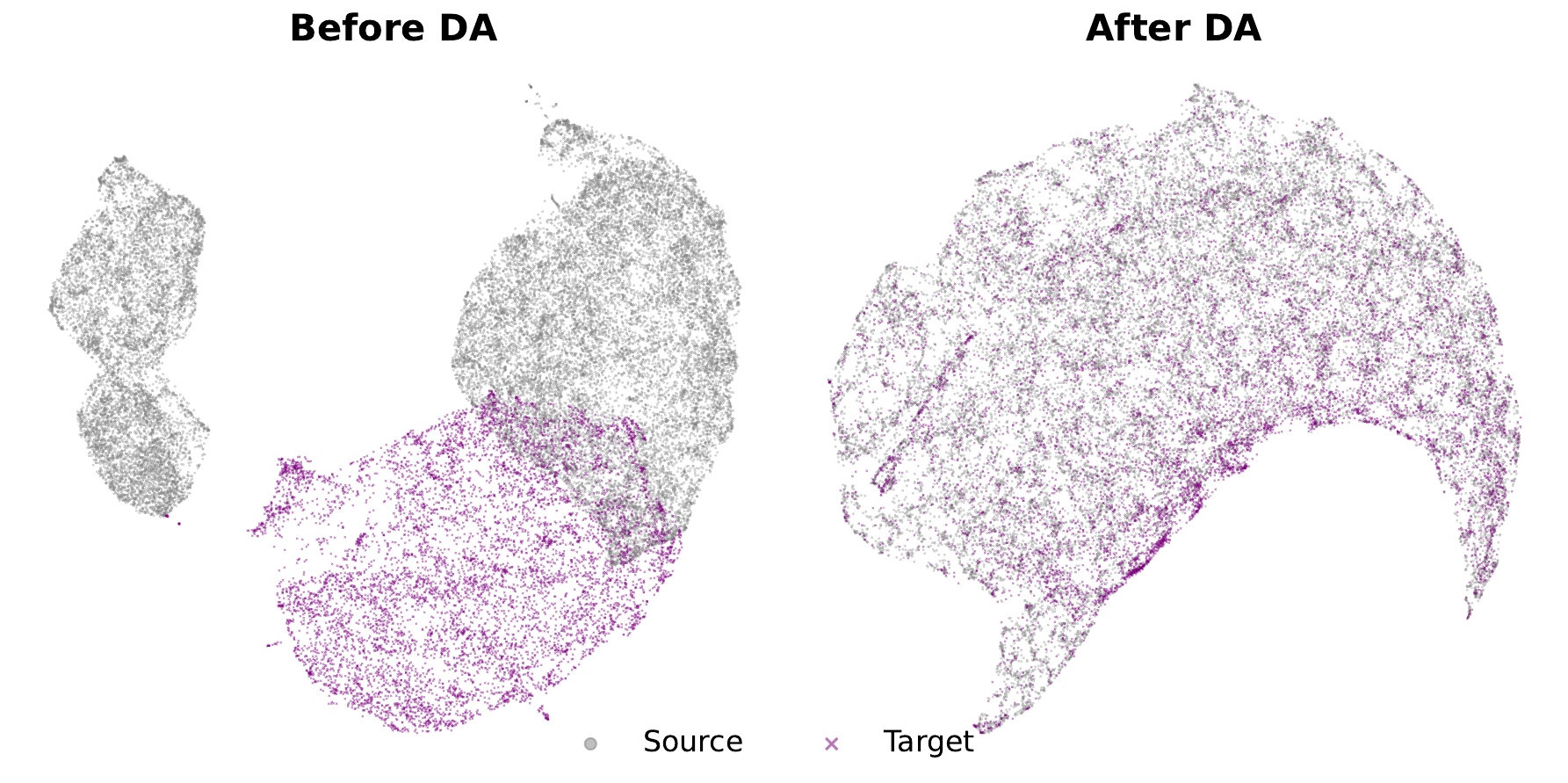}
    \caption{UMAP showing the distribution of source (gray) and target (purple) simulations before DA (left) and after DA (right). Before DA the source and target distributions are highly discrepant. After DA the source and target distributions are indistinguishable suggesting that they are domain-invariant to each other.}
    \label{fig:umap_da}
\end{figure*}

To illustrate the effect of DA, we use Uniform Manifold Approximation and Projection \citep[UMAP;][]{mcinnes_umap_2020} to project the high-dimensional feature space into two dimensions. Figure~\ref{fig:umap_da} shows the feature distributions before (left) and after (right) DA. Without DA, features from different simulations occupy distinct regions in the projected space, indicating a domain shift between them. After applying DA, the features are significantly more aligned, with much greater overlap between the simulations. This alignment suggests that the DA model successfully learns a representation that is less sensitive to simulation-specific effects. %

\section{Results} \label{sec:results}
We now assess the performance of \haloflow~and its domain-adaptive extension, \haloflow$^{\text{DA}}$, in inferring halo masses across different simulation domains. %

\subsection{Domain Shift Degrades Posterior Inference} \label{subsec:domain_shift}

\begin{figure*}[t!]
    \centering
    \includegraphics[width=0.75\linewidth]{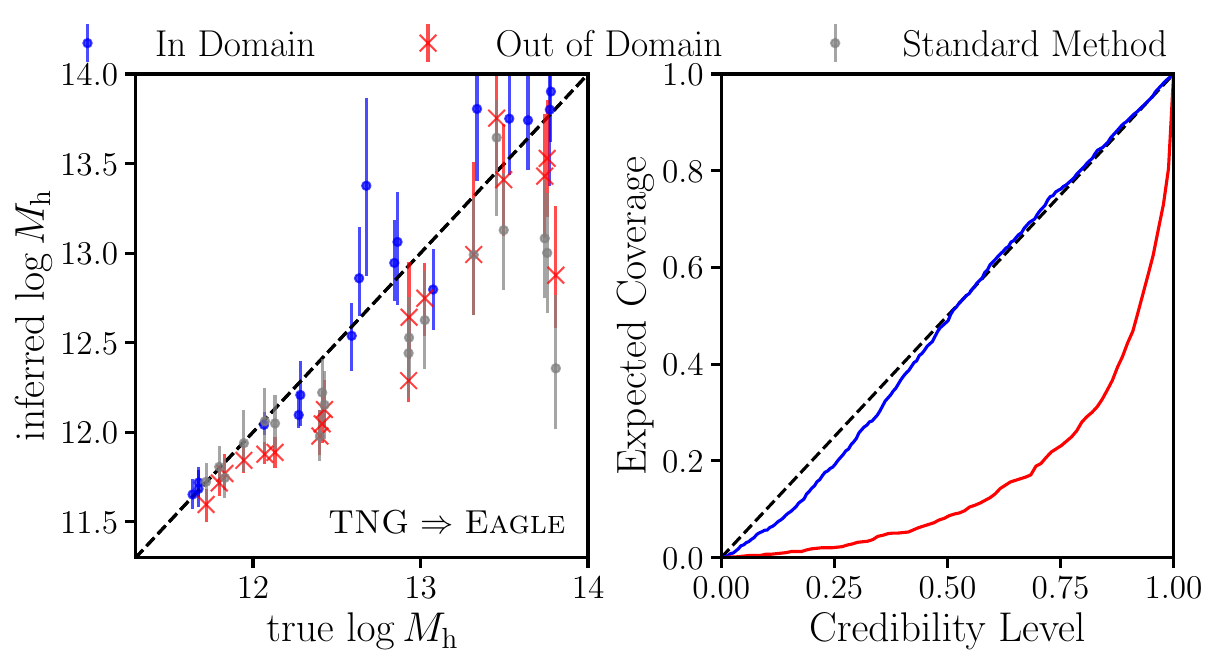}
\caption{\textbf{Left:} Inferred vs.~true \mh~using \haloflow~trained in-domain (blue; \TNG~$\Rightarrow$~\TNG) and out-of domain (red; \TNG~$\Rightarrow$~\Eagle). . Predictions are accurate and well-calibrated in the in-domain case, but show systematic bias under domain shift. 
For reference, we include \mh~constraints using the standard approach based on the SHMR applied out-of-domain (gray). 
The standard approach also exhibits significant systematic bias.
    \textbf{Right:} Coverage plots for the \haloflow~posteriors. 
    In-domain scenario shows near-ideal calibration (black dashed), while out-of-domain shows poorer calibration, indicating overconfident and unreliable posteriors.}
    \label{fig:model_mismatch}
\end{figure*}

Figure~\ref{fig:model_mismatch} illustrates the impact of domain shift on \mh~posteriors. Here, \haloflow~is trained on \TNG, and tested on both \TNG~(in-domain; blue) and \Eagle~(out-of-domain;~red). In the in-domain case, inferred \mh~closely match the true values, with $77\%$ of the halos lie within their quoted $1\sigma$ interval of the posterior, consistent with slightly conservative but well-calibrated posteriors.
However, when \haloflow~is applied to the out-of-domain, the \mh~posteriors become substantially less reliable with only $48 \%$ within their quoted $1\sigma$ interval of the posterior. 
The posteriors are significantly overconfident under the domain shift. %

As additional validation, we perform a ``data to random point” \citep[DRP;][]{lemos_sampling-based_2023} coverage test. For each sample, we evaluate distances between the samples drawn from the two NPEs with a random point in parameter space. We compare this distance to the distance between the true parameter values %
and the random point to derive an estimate of the expected coverage probability.
In the right panel of Figure~\ref{fig:model_mismatch}, we present a quantitative assessment of the posteriors using DRP coverage plots. 
In the in-domain case, the predicted posteriors yield near-ideal coverage (close to the black dashed one-to-one line), implying that \haloflow~provides near optimal estimates of the true posteriors. However, in the out-of-domain case, coverage falls below the ideal diagonal, confirming that the posteriors from the NPE is overconfident and underestimates the uncertainties. 

To more directly quantify inference performance in the context of DA, we define a normalised residual, 
\begin{equation}
    \beta  = \frac{|m - \hat m|}{\sigma_{\hat m}}
\end{equation}
where $\hat m$ and $\sigma_{\hat m}$ are the median and standard deviation of the NPE $M_h$ posteriors and  $m$ is the true \mh.
For an unbiased and perfectly calibrated Gaussian posterior, the median $\beta \simeq 0.67$. 
$\beta \ll 0.67$ indicate  conservative uncertainties, while $\beta \gg 0.67$ signals overconfident or systematically biased posteriors.
For the example in Figure~\ref{fig:model_mismatch}, we find a median $\beta=0.58$ for the in-domain case. 
In contrast, the out-of-domain scenario yields a median $\beta=1.66$.

\begin{figure*}[h!]
    \centering    
    \includegraphics[width=\linewidth]{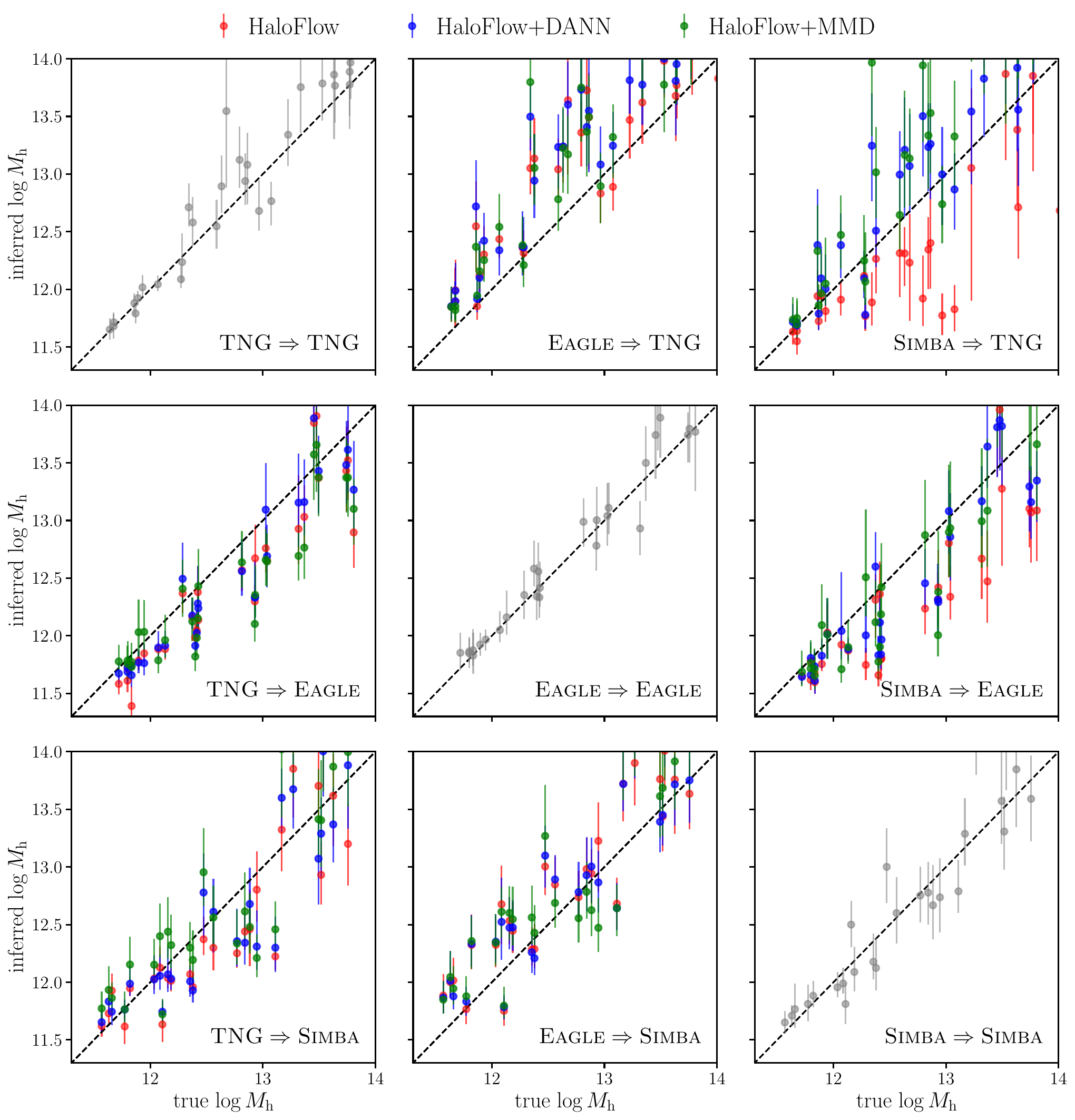}
    \caption{Inferred vs.~true halo masses across all train$\Rightarrow$test simulation pairs. Each panel corresponds to one combination (row: test sim, column: train sim). Scatter points show median inferred $M_h$ with 1$\sigma$ errorbars from: \haloflow~only (red), \haloflow+DANN (blue), and \haloflow+MMD (green). Diagonal panels show in-domain inferences with \haloflow~(gray) for reference. Off-diagonal panels reveal model misspecification for \haloflow~(red) and demonstrate that DANN and MMD (blue and green) improve the robustness of \mh~inference across simulations. }
    \label{fig:sim_pairs_with_da}
    \vspace{-0.5cm}
\end{figure*}

\subsection{Domain Adaptation Improves Generalization}
To test whether DA can improve inference across simulations, we apply both DANN and MMD-based adaptation to all train/test simulation combinations. 
We present our results in Figure~\ref{fig:sim_pairs_with_da}. Each panel corresponds to a specific training (row) $\Rightarrow$ testing (column) simulation. For each setup, we compare posterior medians (with $1\sigma$ error bars) for three NPEs: baseline \haloflow~(red), \haloflow+DANN (blue) and \haloflow+MMD (green).
We only plot a subset of galaxies in the simulations for clarity. 
For reference, we include in the diagonal panels the performance of \haloflow~in-domain (gray).

In off-diagonal cases (out-of-domain), where the test simulation differs from the training simulations, we find systematic deviations for \haloflow~alone--especially in \Simba~$\Rightarrow$ \TNG~and \Simba~$\Rightarrow$~\Eagle, where the median $\beta$ increases to $1.11$ and $1.62$, respectively. Both DA methods significantly reduce these biases and improve calibration where $\beta$ is $0.64$ and $0.70$ respectively.
Moreover, both exhibit adaptive uncertainty behavior: in many cross-domain cases the posterior error bars are widened relative to the in-domain baseline, particularly where the bias and scatter are largest, and remain narrower when the predictions are closer to the one-to-one line. Overall, DA significantly improves the generalization of \haloflow, yielding more robust and better calibrated posteriors across simulations.

To quantitatively assess the effect of DA, we compute the median $\beta$ of \mh~for all train$\Rightarrow$test pairs, shown in Table~\ref{tab:bias_results}. For a baseline reference, in-domain scenarios (diagonal entries), posteriors from the baseline \haloflow~have $\beta \approx 0.5–0.6$. 
In out-of-domain settings, both DANN and MMD reduce $\beta$ in nearly all cases, indicating improved inference quality. We also show the fraction of halos whose masses are within $1\sigma$ of the estimated posteriors for each of the runs in Table~\ref{tab:fractional_unc}.

MMD consistently outperforms DANN, achieving the lowest $\beta$ across all simulation combinations. The most dramatic improvement occurs for the \Simba~$\Rightarrow$~\Eagle~transfer, where $\beta$ drops from 1.62 (no DA) to 0.70 (with MMD) -- a 57\% reduction. This highlights the effectiveness of MMD in aligning feature distributions across physically different simulations.
DANN also yields meaningful improvements, particularly when training on \TNG~or \Simba. However, its performance is less consistent: for instance, when trained on \Eagle, DANN slightly increases $\beta$, suggesting some instability in the adversarial training process.

\begin{figure*}
    \centering
    \includegraphics[width=0.75\linewidth]{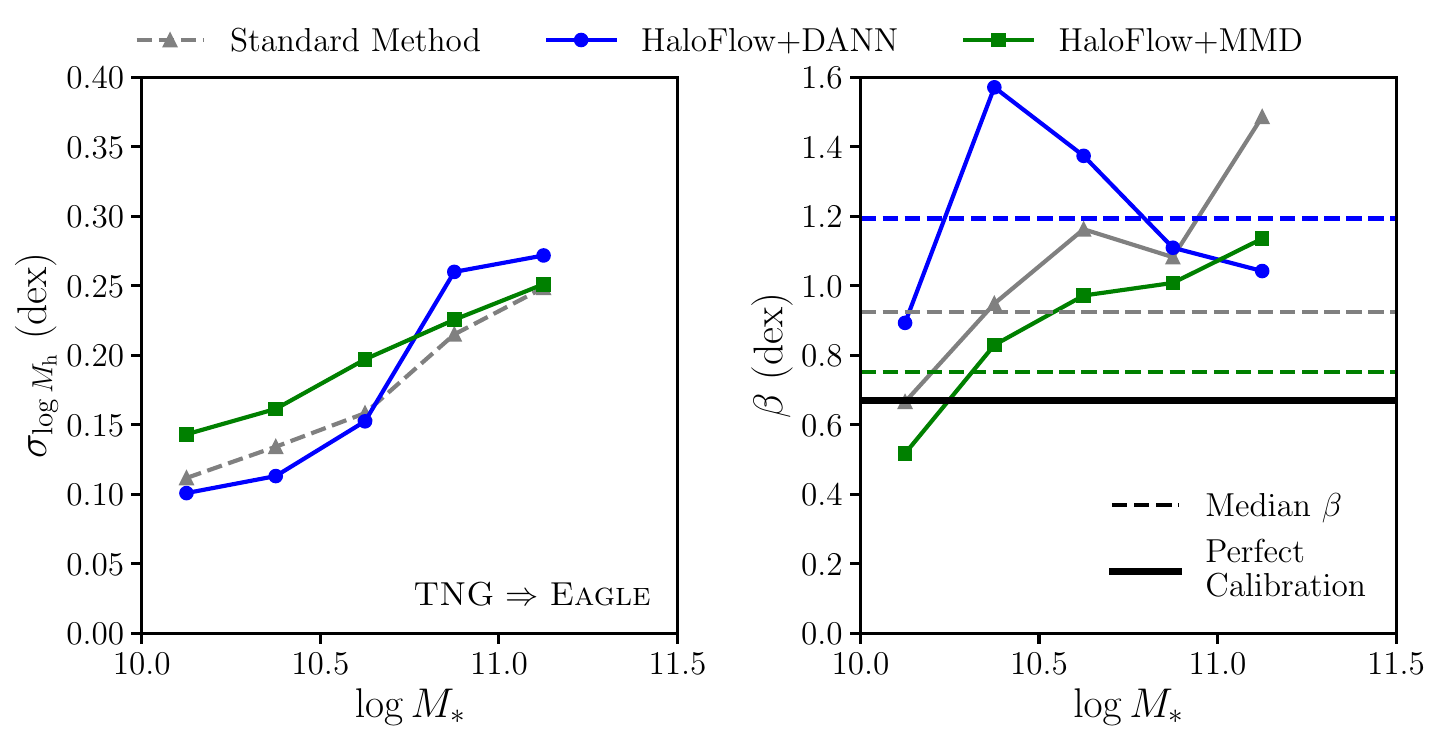}
    \caption{Comparison of the precision (left) and bias (right) of \mh~posteriors inferred using \haloflow$^{\text{DA}}$ (MMD, green; DANN, blue) and the standard method based on the SHMR for the \TNG~$\Rightarrow$~\Eagle~scenario. 
    \textbf{Left:} Median $1\sigma$ of the \mh~posteriors ($\sigma_{\log M_{\rm h}}$), as a function of $M_*$.
    Both the \haloflow$^{\text{DA}}$ methods yield similar precision to that of the standard method even during out-of-domain inference. 
    \textbf{Right:} Median $\beta$ as a function of $M_*$. 
    The dashed lines represent the median $\beta$ for the complete test set. The black bold line is the $\beta \simeq 0.67$ for a perfectly calibrated posterior. 
    The posteriors from \haloflow$^{\text{DA}}$ using MMD is closer to optimally calibrated posteriors than that of the standard method while maintaining comparable precision.}
    \label{fig:precision}
\end{figure*}

To benchmark DA further, we compare it against the standard approach based on the stellar–halo mass relation (SHMR). 
To mimic the standard approach, we use \haloflow~trained on the training simulation to obtain a posterior over stellar mass from photometry in the test simulation. 
This step is analogous to conventional SED modeling. 
Then we draw samples %
$M_*' \sim p(M_*\mid X_{\rm mags})$  and propagate them through the training simulation’s SMHR to infer halo masses. 
In Figure~\ref{fig:precision}, we compare the precision (left) and $\beta$ (right) of the \mh~posteriors derived from \haloflow$^{\rm DA}$ (green, MMD; blue, DANN) versus the standard method (gray dashed) as a function of $M_*$. 
The posteriors are for the \TNG~$\Rightarrow$~\Eagle~scenario. 
We use the standard deviation of the $M_h$ posterior ($\sigma_{\log M_h}$) to represent precision of the posterior. 
\haloflow$^{\rm DA}$ achieves precision comparable to the standard method. 
The \haloflow$^{\rm DA}$ run with MMD also shows improvement in $\beta$ (0.76 compared to 0.94 for the standard method). The horizontal dashed lines represent the overall median $\beta$ for all three runs and the black solid represents an unbiased and perfectly calibrated posterior ($\beta \simeq 0.67$). Overall, \haloflow$^{\rm DA}$ provides a more robust and less biased inference than the standard method, without sacrificing precision.

\begin{table*}[]
\caption{Median $\beta$ of \mh~for all train/test simulation pairs. Rows indicate test simulations, columns indicate train simulations, matching the layout of Figure~\ref{fig:sim_pairs_with_da}. Values are reported as \haloflow~/ \haloflow+DANN / \haloflow+MMD. Diagonal entries correspond to in-domain performance where DA is not applied. Bold values indicate best (lowest) $\beta$ for each pair.}
\label{tab:bias_results}
\centering
\begin{tabular}{lccc}
\toprule
\textbf{Test $\backslash$ Train} & \textbf{\TNG} & \textbf{\Eagle} & \textbf{\Simba} \\
\midrule
\textbf{\TNG} 
& \textbf{0.58} / — / —  & 1.26 / 1.38 / \textbf{1.10}  & 1.11 / 0.89 / \textbf{0.64} \\
\textbf{\Eagle} & 1.66 / 1.18 / \textbf{0.76} & \textbf{0.54} / — / — & 1.62 / 1.06 / \textbf{0.70} \\
\textbf{\Simba} & 1.12 / 0.95 / \textbf{0.94} & 1.37 / 1.50 / \textbf{1.32} & \textbf{0.58} / — / — \\
\bottomrule
\end{tabular}
\end{table*}

\begin{table*}[]
\caption{Fraction of halos within the quoted $1\sigma$ interval of the posterior. Layout of this table matches Table \ref{tab:bias_results}. Bold values indicate the highest fraction for each pair.}
\label{tab:fractional_unc}
\centering
\begin{tabular}{lccc}
\toprule
\textbf{Test $\backslash$ Train} & \textbf{\TNG} & \textbf{\Eagle} & \textbf{\Simba} \\
\midrule
\textbf{\TNG} 
& \textbf{0.75} / — / —  & 0.35 / 0.31 / \textbf{0.46}  & 0.46 / 0.58 / \textbf{0.70} \\
\textbf{\Eagle} & 0.31 / 0.44 / \textbf{0.60} & \textbf{0.80} / — / — & 0.32 / 0.48 / \textbf{0.64} \\
\textbf{\Simba} & 0.45 / 0.52 / \textbf{0.53} & 0.34 / 0.30 / \textbf{0.37} & \textbf{0.76} / — / — \\
\bottomrule
\end{tabular}
\end{table*}

\section{Discussion} \label{sec:discuss}
We investigate how domain shifts among different simulations impact \mh~inference when using \haloflow. 
Having established that DA improves the robustness of \mh~posteriors out-of-domain, we next examine (i) the role of the DA loss weight $\lambda$, (ii) where DA should be applied within the network, (iii) the resulting posterior calibration and uncertainties, (iv) implications for information-maximizing compression methods more broadly, and (v) fundamental limitations of DA across physically distinct models.

\subsection{Sensitivity to the DA loss $\lambda$}
The parameter $\lambda$ in Eq.~\ref{eq:loss} plays a crucial role in balancing the trade-off between enforcing domain invariance and preserving information relevant for inferring \mh. 
To probe this trade-off, we varied $\lambda$ for some of the worst-performing out-of-domain transfers (highest median $\beta$ in Table~\ref{tab:bias_results}).%

When $\lambda$ is large ($\gtrsim 1$), the DA network places more emphasis on aligning feature distributions across domains, effectively minimizing differences in the simulations. %
While this improves robustness to domain shift, it degrades feature extraction by discarding features that are informative for \mh, leading to broader posteriors. Conversely, when $\lambda$ is small ($< 1$), the network prioritizes minimizing the regression loss, retaining more \mh-relevant information but becoming more sensitive to discrepancies between domains. In the limit $\lambda = 0$, the DA objective vanishes and the network performs a lossy compression of the input features $\mathbf{X}$ without any domain adaptation; the method reduces to standard \haloflow~with data compression.

In practice, we find that the behavior of $\lambda$ depends strongly on the choice of DA method. 
For DANN, tuning $\lambda$ proved challenging due to inherent instability of adversarial training. As $\lambda$ increases, the domain classifier becomes increasingly dominant, often overpowering the feature extraction. %
While we followed the scheduling for $\lambda$ from \cite{csurka_domain-adversarial_2017}, ramping up $\lambda$ gradually (see Eq.~\ref{eq:lambda_dann}), we observed substantial variability in the performance based on the choice of training and test domains. In some cases, this led to higher median $\beta$ values than the baseline \haloflow~when applied to out-of-domain simulations, particularly for runs in \Eagle~$\Rightarrow$~\TNG~or~\Simba. 

In contrast, the MMD-based method was substantially more stable across a wide range of $\lambda$ values. After testing several configurations, we found that a fixed $\lambda = 0.5$ consistently yielded reliable performance across all simulation pairs. This suggests that the kernel-based approach of MMD, which avoids adversarial dynamics entirely, offers a more robust solution in data-limited regimes or when simulation differences are pronounced (more details in Appendix~\ref{app:train}).

\subsection{Where to apply DA?}
While DA can mitigate discrepancies between domains, we found that its success highly depends on \textit{where} and \textit{how} the compressed features for \haloflow$^{\rm DA}$ are defined, consistent with previous work \citep{zhao2019learning, sicilia_domain_2023}. Early in our experiments, we attempted to enforce domain invariance on a highly compressed representation, using the two-dimensional regressor output, $\hat{\mathbf{\theta}}$. 
For DANN, this configuration performed poorly, producing larger bias $\beta$ values than the baseline. A larger value of $\beta$ does not necessarily imply a narrower posterior; instead it means that the absolute error increased relative to the inferred uncertainty. This likely occurred because enforcing strict domain invariance at such a low-dimensional stage discarded information essential for accurate inference: by forcing $\hat{\mathbf{\theta}}$ to appear similar across simulations, the network lost flexibility to account for subtle but physically meaningful variations in galaxy properties. 

Interestingly, the same approach using MMD performed substantially better. Even with this extreme compression, the distribution-level alignment on $\hat{\mathbf{\theta}}$ preserved enough $\theta$-informative structure to yield stable and relatively accurate posterior estimates. This contrast suggests that the alignment, which relies on a classifier-based criterion, is highly sensitive to the dimensionality of the features, whereas the distribution-matching approach is more robust to where alignment occurs within the processing pipeline.

Our final DA inference uses the compressed DA features differently depending on the method. For DANN, we perform DA at the level of the higher-dimensional embedding produced by the feature extractor, which encodes galaxy morphology, colors, and other parameters that remain informative for \mh~estimation. This richer representation allows learning of domain-invariant yet physically meaningful features, improving stability and posterior accuracy. By contrast, the MMD approach achieves effective DA when applied to the regressor output, aligning simulations in the space of predicted quantities without substantial information loss.

\subsection{Posterior calibration, uncertainity and robustness}
A useful property of both DA methods is that neither increases posterior uncertainty arbitrarily. In fact, both approaches produce uncertainty estimates that are adaptively narrower when the inferred \mh~is accurate and wider in ambiguous or poorly calibrated regions.
This behavior is consistent with our training objective: the DA losses act only on the internal feature representations, while the posterior over \mh~is still constrained for simulated pairs $(\mathbf{X}, \theta)$. Where the target-domain data resemble the training simulations, DA primarily removes domain-dependent shifts without reducing the information content, so the posteriors remain tight but better calibrated. In regions where the simulations disagree more strongly or the mapping between galaxy properties and \mh~is less well constrained, DA prevents the network from extrapolating aggressively, leading to broader, more conservative posteriors.

\subsection{Implications for information-maximizing compression}
These insights extend beyond the specific methods discussed here. Neural networks are commonly used in astronomy and cosmology for compressing data into informative summaries for inference, such as Information Maximising Neural Networks \citep[IMNNs; \eg][]{charnock_automatic_2018, makinen_lossless_2021, lucas_makinen_hybrid_2025}, Moment Networks \citep{jeffrey_solving_2020}, %
or convolutional neural networks (CNNs) applied directly to maps and images \citep[\eg][]{hahn_forward_2023, eisert_ergo-ml_2024}.
These methods all face the same challenges as \haloflow~when training and testing data differ due to domain shifts. %

Without explicit DA, their summaries can fail to capture all of the parameter information present in the data once the domain shifts, leading to degraded inference. Moreover, compressing to very low-dimensional summaries can make this problem worse and can also make DA more susceptible. As discussed above, our experiments with DANN on a compressed representation yielded more biased inference the baseline \haloflow. Kernel-based approaches such as MMD mitigated these issues and often improved robustness in our tests, but they do not \textit{eliminate} simulation mismatch entirely. The broader takeaway is that studies using %
compressors should explicitly assess how domain shifts affect their compressed features and the resulting posteriors, and cannot assume that a summary that works well on the training simulations will remain reliable out-of-domain.

\subsection{Limitations of DA} %
Although \haloflow$^{\rm DA}$ improves the overall robustness of the \mh~inference, its performance still depends on the degree of physical mismatch between simulations. For instance, \Simba, implements more aggressive AGN feedback than either \TNG~or \Eagle; in \Simba, this leads to earlier quenching and alters their stellar mass and gas content at fixed \mh \citep{ayromlou_feedback_2023, wright_baryon_2024}, making \Simba~a particularly challenging target. 
When training on \Eagle\ and testing on \Simba, the baseline \haloflow\ yields a median $\beta = 1.37$, higher than for most other simulation pairs.
DA does not significantly improve the robustness ($\beta_{\rm MMD}=1.32$), which suggests that 
DA cannot fully account for divergences in the underlying physical models. %

Even if DA brings the feature distributions closer together, it cannot fully remove the mismatch because the underlying galaxy–halo connection itself is different. In practice, DA can shift and reweight the existing distributions so that simulations look more alike, but it cannot generate physical behavior that is absent from the training simulations.
This also implies that DA only works well if the training data used for DA encompasses that of the test set.
In future work, when we apply \haloflow$^{\rm DA}$ to observations, we are assuming that it lies somewhere in between of all the training simulations.

There are further caveats. %
Our results are based on only three hydrodynamical simulation suites: \TNG, \Eagle, and \Simba. While these represent a range of subgrid physics, they do not span the full space of plausible galaxy formation models. This limits our ability to fully assess how DA performs across more extreme physical priors. Expanding the training and testing set to include additional simulations—especially those with varied feedback implementations or alternative dark matter models—will be important for testing the generality of our approach.

This study focuses exclusively on central galaxies with stellar masses above $M_* > 10^{10} M_\odot$ at $z=0.1$. %
This selection limits our conclusions. Lower-mass galaxies, satellites, or higher-redshift populations could exhibit stronger or weaker domain shifts. 
Our results should therefore be interpreted as demonstration on %
a specific galaxy population and direct future applications of it can only be applied to consistent galaxy populations. 
We, however, emphasize that the \haloflow$^{\rm DA}$ can be easily extended by incorporating other galaxy populations in the training. 

Despite these caveats, the improvements shown by \haloflow$^{\rm DA}$ suggest that these techniques will be valuable when applying simulation-trained models to real galaxy observations. Domain shift is unavoidable in observational settings due to differences in measurement noise, selection effects, and the fact that no simulation perfectly captures the physics of galaxy formation. While our experiments are limited to a small set of hydro simulations, the consistent gains from MMD-based adaptation indicate that even approximate alignment of latent feature distributions can improve out-of-domain performance. This opens the possibility of applying \haloflow$^{\rm DA}$ to real HSC-SSP observations.

\section{Conclusions} \label{sec:conclusion}
In this work, we introduced \haloflow$^{\rm DA}$, a domain-adapted extension of the \haloflow~framework that combines SBI with DA to improve the robustness of \mh~inference from galaxy photometry and morphology. Using forward-modeled HSC-like galaxy images from three state-of-the-art hydrodynamical simulations, \TNG, \Eagle, and \Simba, we showed that the standard \haloflow~\mh~posteriors are robust when we train and test on the same simulations but become biased and overconfident when applied across simulations.
This degradation arises from domain shift: each simulation encodes a different galaxy formation model and, thus, induces a different mapping between galaxy observables and \mh.

To address this, \haloflow$^{\rm DA}$~learns domain-adapted representations of galaxy observables that retain the information relevant for \mh~inference while suppressing features specific to individual simulations. We implemented two DA strategies: DANN and MMD. DANN and MMD both aim to align these representations. DANN does this adversarially by training a feature extractor to ``fool'' a domain classifier, while MMD minimizes a statistical distance between the train and test domains' feature distributions directly.
Once the DA network is trained, its domain-adapted representations $c{\bf X}$ are input into the 
\haloflow~SBI framework to produce $M_*$ and \mh~posteriors.%

To quantify the performance of \haloflow$^{\rm DA}$, we apply it across all train$\Rightarrow$test simulation pair combinations of TNG, EAGLE, and SIMBA.  %
Our main findings are:
\begin{itemize}
    \item \textbf{Domain shift significantly degrades \mh ~posteriors, but DA substantially improves robustness}.
     In–domain applications of \haloflow~yield well–behaved \mh~posteriors with median $\beta \simeq 0.5$–0.6 and near–ideal coverage. 
     Out-of-domain, median $\beta\lesssim 1-1.5$, indicating biased and overconfident posteriors.
     \haloflow$^{\rm DA}$ systematically reduces these failures: both DANN and MMD lower $\beta$ in nearly all test cases. %

    \item \textbf{The architecture of the DA matters; kernel–based MMD is particularly robust.}  
    The impact of DA depends on both the method and the representation on which alignment is enforced. DANN, which uses adversarial training, is sensitive to the strength of the DA loss and to the dimensionality of the aligned features. %
    MMD, by contrast, is substantially more robust: a simple configuration with fixed $\lambda = 0.5$ and RBF kernel bandwidth $\sigma = 1.0$ consistently reduces $\beta$ across all train$\Rightarrow$test combinations and achieves the best overall performance. %

    \item \textbf{DA improves robustness but cannot fully compensate for physical mismatches.}%
    ~The effectiveness of \haloflow$^{\rm DA}$ is limited by differences in the underlying galaxy–halo connection. \Simba, for example, has more aggressive AGN feedback than \TNG~and \Eagle, which makes it a more challenging test domain. 
    In this case, DA improves $\beta$, but not comparable to in–domain performance. DA can align observable feature distributions and reduce systematic biases, but it cannot generate physical behavior absent from the training simulations. %
    Thus, the success of \haloflow$^{\rm DA}$ ultimately depends on training over a set of simulations that plausibly bracket the real Universe.
\end{itemize}
These results demonstrate that DA is necessary in making simulation–based \mh~inference robust to the inevitable mismatches between simulations and reality. Although our experiments are restricted to three hydrodynamical suites and to central galaxies with $M_* > 10^{10},M_\odot$ at $z=0.1$, they provide a concrete demonstration: %
even approximate alignment of latent feature distributions via MMD–based DA leads to more accurate, better–calibrated posteriors in out–of–domain settings.

In future work, we will test %
the \haloflow$^{\rm DA}$ \mh~constraints of observed galaxies and groups in HSC-SSP against independent mass constraints from galaxy-galaxy weak lensing measurements \citep[\eg][]{rana_subaru_2022}. We will also compare them to dynamical mass estimates of groups identified in GAMA \citep{driver_empirical_2022}. Once validated, we will apply \haloflow$^{\rm DA}$ to intervening halos observed by the FLIMFLAM spectroscopic survey \citep{lee_constraining_2022}, which targets FRB foreground fields, in order to improve constraints on the CGM baryonic fraction in those systems. 

\appendix
\section{Correction of Halo Mass Prior} \label{app:prior_correction}

The trained \haloflow$^{\rm DA}$ NPE, $q_\phi(\theta \mid \mathbf{X})$, inherits an implicit prior over parameters, $p_{\rm imp}(\theta)$, set by the distribution of the training set~\citep{hahn_haloflow_2024}. Because the training galaxies are distributed according to the underlying stellar- and halo-mass functions, $p_{\rm imp}(\theta)$ reflects these functions.
In this work, we seek to adopt a prior uniform in $\log M_*$ and $\log M_{\rm h}$. 

Following the maximum-entropy prior framework of \citet{handley_maximum-entropy_2019}, we reweight posterior samples to enforce a user-defined target prior $r(\theta)$. For posterior samples $\theta’ \sim q_\phi(\theta \mid \mathbf{X})$, we assign importance weights
\begin{equation}
w(\theta’) = \frac{r(\theta’)}{p_{\rm imp}(\theta’)} ,
\end{equation}
which reduces to
\begin{equation}
w(\theta’) \propto \frac{1}{p_{\rm imp}(\theta’)}
\end{equation}
for a uniform target prior. 

In practice, this correction is implemented via importance sampling and can become unstable when $p_{\rm imp}(\theta)$ is poorly sampled, particularly in the high-mass regime of the training distribution where the stellar- and halo-mass functions are sparsely populated.
This instability is driven by two effects: (i) numerical noise in the estimate of $p_{\rm imp}$ in the high-mass tail, where the halo mass function is sparsely sampled, and (ii) increased uncertainty of the NPE 
in the tails of the posterior. Because the correction scales as $1/p_{\rm imp}$, both effects can produce a small number of samples with disproportionately large weights, causing the reweighted posterior to be dominated by rare tail draws.

We stabilise the implicit-prior correction in two ways. First, to reduce stochasticity in the high-mass tail of the halo-mass distribution, we replace the empirical estimate of the implicit prior with a smooth analytical approximation by fitting a single Schechter function to the training-set halo-mass distribution and evaluating $p_{\rm imp}$ from this fit. Second, to limit sensitivity to inaccuracies at the tails of the NPE, we truncate the posterior samples before applying the improtance weights by discarding draws outside the central $99\%$ interval in \mh~(i.e.~below the $0.5$th percentile or above the $99.5$th percentile of the sampled marginal distributions). Together, these steps substantially improve the numerical stability of the importance sampled posteriors. %
Below, we describe these two steps in further detail. 

\subsection{Fitting the intrinsic halo mass function}

To reduce weight instabilities caused by noise in the empirical estimate of $p_{\rm imp}$ in the high-mass tail, we model the training-set halo-mass distribution with a single-Schechter function. In linear halo mass, the model is

\begin{equation}
\phi(M_h) =
\phi_M
\left(\frac{M_h}{M_h^{\rm char}}\right)^{\alpha}
\exp\left(-\frac{M_h}{M_h^{\rm char}}\right),
\end{equation}
where $\phi_M$ is the normalization, $M_h^{\rm char}$ is the characteristic halo mass, and $\alpha$ is the low-mass slope.
Since inference is performed in logarithmic halo mass,
\begin{equation}
g \equiv \log_{10} M_h,
\end{equation}
we work with the corresponding form per unit $\log_{10} M_h$,

\begin{equation}
\phi(g)
=
\ln(10)\,\phi_M\,
10^{(g - \log_{10} M_h^{\rm char})(\alpha + 1)}
\exp\left[-10^{(g - \log_{10} M_h^{\rm char})}\right].
\label{eq:log_schehter}
\end{equation}

We bin the training-set halo masses in $g$ with bin edges $\{g_i\}$ and observed counts $y_i$. We fit the Schechter parameters $\boldsymbol{\psi}=(\phi_M,\log_{10} M_h^{\rm char},\alpha)$ by maximizing the Poisson likelihood of the binned counts. The expected number of halos in bin $i$ is
\begin{equation}
\mu_i(\boldsymbol{\psi})=\int_{g_i}^{g_{i+1}}\phi(g\mid\boldsymbol{\psi})\,\mathrm{d}g,
\end{equation}
and, dropping terms constant in $\boldsymbol{\psi}$, the negative log-likelihood minimized during fitting is
\begin{equation}
\mathcal{N}(\boldsymbol{\psi})=
\sum_i\left[\mu_i(\boldsymbol{\psi})-y_i\ln\mu_i(\boldsymbol{\psi})\right].
\end{equation}

\subsection{Regularized halo-mass importance weights}

For a uniform target prior, the maximum-entropy prior correction implies importance weights $w(\theta') \propto 1/p_{\rm imp}(\theta')$. We approximate the implicit prior over $g$ using the fitted Schechter model evaluated at ten equally spaced bins within the range of the halo-mass function.

The maximum-likelihood parameters $\boldsymbol{\psi}_{\rm MLE}$ define a smooth parametric estimate of the training-set halo-mass distribution. The corresponding expected counts per bin,
\begin{equation}
\mu_i^{\rm prior} \equiv \mu_i(\boldsymbol{\psi}_{\rm MLE}),
\end{equation}
define a discrete binned prior mass
\begin{equation}
p_i^{\rm prior} \equiv \frac{\mu_i^{\rm prior}}{\sum_k \mu_k^{\rm prior}}.
\end{equation}
For a posterior sample $\theta^{(j)}$ with halo mass $g^{(j)}$, we identify the corresponding mass bin $i(j)$ and assign an importance weight
\begin{equation}
w_h^{(j)} \propto \frac{1}{p_{i(j)}^{\rm prior}}.
\end{equation}

To ensure numerical stability, halo masses are clipped to the fitted mass range prior to bin assignment, and a small positive floor is imposed on $p_i^{\rm prior}$. Finally, for convenience, the weights are renormalized such that
\begin{equation}
\sum_j w_h^{(j)} = N_{\rm samp},
\end{equation}
which preserves relative weighting and does not affect posterior expectations.

\subsection{Truncation of NPE posterior tails}

In addition to the scarcity of high-mass halos in the training data, a second source of numerical instability arises from increased uncertainty in the tails of the normalizing flow posteriors. While the NPEs provide accurate posterior estimates in regions well supported by the training data, the extreme tails of the posterior distributions are often less well-constrained. When combined with importance weighting, these poorly constrained regions can %
lead to inflated variances and unstable posterior summaries.

To mitigate this effect, we truncate the NPE posterior samples by removing the most uncertain tail regions \emph{before} the importance weighting. Specifically, for each inferred posterior we compute the $0.5$th and $99.5$th percentiles of the marginal halo-mass posterior samples, denoted $M_h^{0.5}$ and $M_h^{99.5}$, and discard draws with $M_h < M_h^{0.5}$ or $M_h > M_h^{99.5}$. This truncation removes the region where we suspect the NPEs are least reliable, while preserving the central $99\%$ of the posterior mass relevant for inference.

\begin{figure*}[h!]
    \centering
    \includegraphics[width=0.5\linewidth]{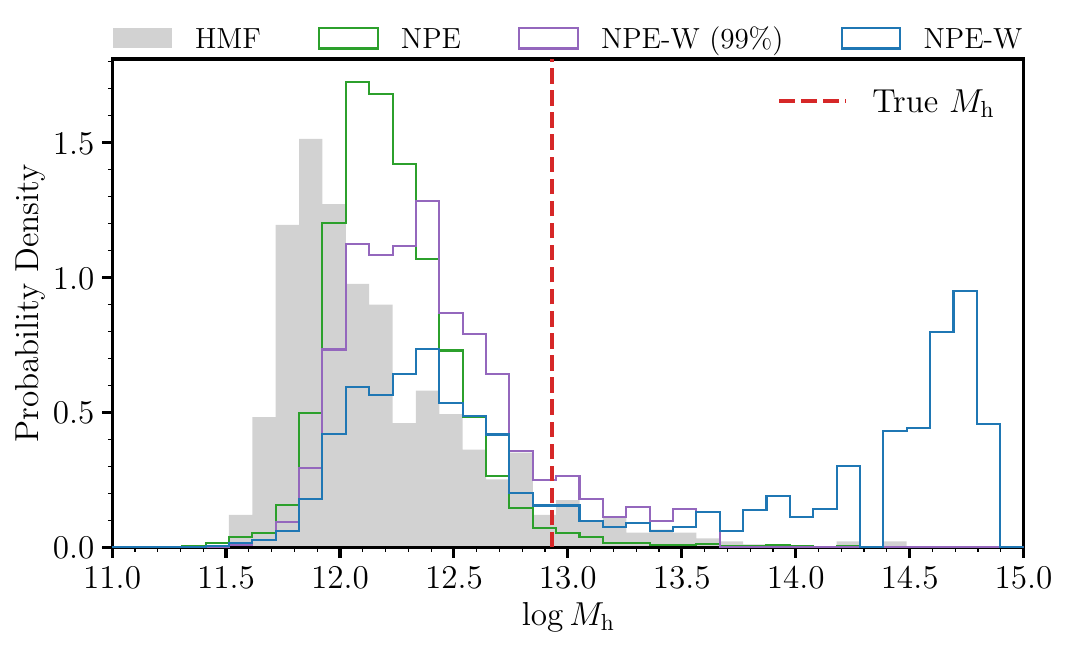}
    \caption{
Comparison of halo-mass posterior distributions inferred using different methods.
The gray histogram shows the intrinsic halo mass function (HMF) of the training data.
The green curve corresponds to the unweighted NPE posterior.
The blue curve shows the importance-weighted NPE posterior without truncation, which exhibits an amplified and unstable high-mass tail.
The purple curve shows the importance-weighted NPE posterior after truncation at the central $99^{\rm th}$ percentile, yielding a stable distribution.
The vertical red dashed line marks the true halo mass, $M_h$.
}
    \label{fig:mh_posteriors}
\end{figure*}

We emphasize that this truncation is applied before the importance weighting and does not introduce an explicit mass-dependent prior. Instead, it serves as a regularization step that limits the influence of regions where the posterior is poorly constrained. %

Figure~\ref{fig:mh_posteriors} illustrates the effect of importance weighting and posterior truncation on the inferred halo-mass distributions. The gray filled histogram shows the probability density of the intrinsic halo mass function of the training data. The green histogram corresponds to the unweighted \haloflow$^{\rm DA}$ NPE. The blue histogram corresponds to the importance-weighted NPE without any truncation. This histogram is amplified and unstable in the high-mass end. The purple curve shows the importance-weighted NPE with truncation at central $99^{\rm th}$ percentile where the distribution is much more stable in the high-mass tail. The red vertical dashed line represents the true \mh.

\section{Construction of DA Sample Weights} \label{app:weights}

In Eq.~\ref{eq:reg_loss} we introduced per-sample weights $w_i$ in the regression loss,
\begin{equation}
\ml_{\text{reg}} = \sum_{i=1}^N w_i\, ||\hat\theta_i - \theta_i||^2
\end{equation}
in order to (i) balance the relative contribution of different simulations (domains) and (ii) mitigate the strong imbalance in halo-mass sampling. Each weight is defined as the product of a domain-balancing factor and a mass-dependent factor,
\begin{equation}
w_i = w_{\text{dom}}(d_i)\,w_{\text{mass}}(M_{*,i}),
\end{equation}
where $d_i$ denotes the simulation (domain) of sample $i$ and $M_{,i}$ is its stellar mass.

\subsection{Domain Balancing}

Let $d \in {1,\dots,K}$ index the simulations and $N_d$ be the number of training samples from domain $d$. Without correction, domains with more samples would dominate the regression loss. To ensure that each domain has approximately equal influence on the loss, we assign each galaxy from domain $d$ a domain weight
\begin{equation}
w_{\text{dom}}(d) \propto \frac{1}{N_d}.
\end{equation}
In practice, this is implemented by constructing an array of domain sizes and computing
\begin{equation}
w_{\text{dom}}(d_i) = \frac{S}{N_{d_i}},
\end{equation}
where $S$ is the sum of the distinct domain sizes. No additional global normalization is applied; the overall scale of $w_{\text{dom}}$ is absorbed into the effective learning rate.

\subsection{Mass-Dependent Weighting}

Even within a given domain, the training-set is strongly imbalanced because it inherits the underlying halo-mass function, with low-mass halos greatly outnumbering high-mass halos. Without additional weighting, the loss in Eq.~\ref{eq:reg_loss} would be dominated at the low-mass regime. To mitigate this imbalance, we construct a smooth parametric approximation to training-set mass distribution and up-weight underrepresented masses. %

We model the one-dimensional training-set distribution in log-mass using the Schechter form in log space, $\phi(m)$, defined in Appendix~A (Eq.~\ref{eq:log_schehter}). We fit the Schechter parameters to the combined training set by (i) forming a histogram of $m$ using 20 uniform bins spanning the training-set range, (ii) converting bin counts to a density per dex (including the simulation volume normalization), and (iii) performing a non-linear least-squares fit to the binned density evaluated at the bin centers.

Given the fitted $\phi(m)$, we compute an unnormalized per-sample mass weight as the inverse of the fitted density evaluated at each sample,
\begin{equation}
\tilde{w}_{\text{mass}}(M{,i}) = \frac{1}{\phi(M_{,i})}.
\end{equation}
We then rescale it by the minimum value across the training set so that the smallest mass weight is unity:
\begin{equation}
w_{\text{mass}}(M_{,i}) = \frac{\tilde{w}_{\text{mass}}(M{,i})}{\min_j \tilde{w}_{\text{mass}}(M{,j})}.
\end{equation}
and clip the mass-dependent weights from above to avoid excessively large values,
\begin{equation}
w_{\text{mass}}(M_{,i}) = \min\bigl(w_{\text{mass}}(M_{*,i}), 10^2\bigr),
\end{equation}
so that $w_{\text{mass}} \in [1, 10^2]$. No further normalization of $w_{\text{mass}}$ is applied.
The final per-sample weights used in Eq.~\ref{eq:reg_loss} are then given by
\begin{equation}
w_i = w_{\text{dom}}(d_i),w_{\text{mass}}(M_{*,i}),
\end{equation}
which up-weights underrepresented domains and the high-mass tail while down-weighting the most heavily populated regions of parameter space.

\section{Network Architectures and Training Hyperparameters} \label{app:arch}

\subsection{DANN Architecture}

The DANN model consists of three fully connected modules: a feature extractor that maps the input feature vector $\mathbf{X}$ to a 32-dimensional latent representation, a label-predictor (regressor) that outputs $(M_*, M_{\rm h})$, and a domain classifier that predicts the simulation label. All hidden layers use SiLU activations. In the \haloflow$^{\rm DA}$ pipeline, the 32-dimensional latent output of the feature extractor is used as the compressed representation $c\mathbf{X}$.

\begin{table}[h]
    \centering
    \caption{DANN architecture. All hidden layers use SiLU activations.}
    \label{tab:dann_arch}
    \begin{tabular}{llll}
        \toprule
        Block & Layer & Output dim. & Activation \\
        \midrule
        \multirow{3}{*}{Feature extractor} 
          & Linear($d_{\rm in}$, 128) & 128 & SiLU \\
          & Linear(128, 64)           & 64  & SiLU \\
          & Linear(64, 32)            & 32  & SiLU \\
        \midrule
        \multirow{3}{*}{Label predictor} 
          & Linear(32, 16) & 16 & SiLU \\
          & Linear(16, 8)  & 8  & SiLU \\
          & Linear(8, 2)   & 2  & ---  \\
        \midrule
        \multirow{3}{*}{Domain classifier} 
          & Linear(32, 16) & 16 & SiLU \\
          & Linear(16, 8)  & 8  & SiLU \\
          & Linear(8, 4)   & 4  & ---  \\
        \bottomrule
    \end{tabular}
    \tablecomments{The two outputs of the label predictor correspond to $(M_*, M_{\rm h})$. The four outputs of the domain classifier correspond to the simulations. The 32-dimensional latent features from the feature extractor are used as $c\mathbf{X}$.}
\end{table}

\subsection{MMD Architecture} \label{app:mmd_arch}

The MMD model shares the same high-level structure as DANN but omits the domain classifier. It contains a feature extractor that maps $\mathbf{X}$ to a 32-dimensional latent representation and a regressor that outputs $(M_*, M_{\rm h})$. The MMD loss is computed on the 32-dimensional latent features, while the 2-dimensional regressor output is used as $c\mathbf{X}$ in \haloflow$^{\rm DA}$.

\begin{table}[h]
    \centering
    \caption{MMD architecture. All hidden layers use SiLU activations.}
    \label{tab:mmd_arch}
    \begin{tabular}{llll}
        \toprule
        Block & Layer & Output dim. & Activation \\
        \midrule
        \multirow{4}{*}{Feature extractor} 
          & Linear($d_{\rm in}$, 256) & 256 & SiLU \\
          & Linear(256, 128)          & 128 & SiLU \\
          & Linear(128, 64)           & 64  & SiLU \\
          & Linear(64, 32)            & 32  & ---  \\
        \midrule
        \multirow{3}{*}{Label predictor} 
          & Linear(32, 16) & 16 & SiLU \\
          & Linear(16, 8)  & 8  & SiLU \\
          & Linear(8, 2)   & 2  & ---  \\
        \bottomrule
    \end{tabular}
    \tablecomments{The two outputs of the label predictor correspond to $(M_*, M_{\rm h})$. The 32-dimensional latent features from the feature extractor are used to compute the MMD loss (Eq.~\ref{eq:loss_mmd}), while the 2-dimensional output of the label predictor is used as the compressed representation $c\mathbf{X}$ for the MMD runs.}
\end{table}

\subsection{Training Hyperparameters} \label{app:train}

Table~\ref{tab:train_hparams} summarizes the main training hyperparameters used for the DANN and MMD runs. Unless otherwise noted, all other optimizer settings are left at their PyTorch defaults.

\begin{table}[h]
    \centering
    \caption{Training hyperparameters for the DANN and MMD models.}
    \label{tab:train_hparams}
    \begin{tabular}{lll}
        \toprule
        Parameter & DANN & MMD \\
        \midrule
        Optimizer 
            & AdamW \citep{loshchilov_decoupled_2019} 
            & AdamW \\
        Learning rate 
            & $10^{-2}$ 
            & $10^{-4}$ \\
        Weight decay 
            & $10^{-5}$ 
            & $5\times10^{-4}$ \\
        Max epochs 
            & 1000 
            & 350 \\
        Batching 
            & full-batch (entire training set) 
            & mini-batch, size 512 \\
        LR scheduler 
            & ReduceLROnPlateau on val. loss \\
            & (factor 0.5, patience 45) 
            & none \\
        Gradient clipping 
            & global norm $\le 1.0$ 
            & none \\
        Early stopping patience 
            & 50 epochs 
            & 10 epochs \\
        DA weight $\lambda$ 
            & schedule $\lambda(p)$ (Eq.~\ref{eq:lambda_dann}) 
            & fixed $\lambda = 0.5$ \\
        Kernel bandwidth $\sigma$ 
            & --- 
            & fixed $\sigma = 1.0$ \\
        \bottomrule
    \end{tabular}
\end{table}

For DANN, we perform full-batch gradient updates: at each epoch the loss is evaluated over the entire training set before performing a single optimizer step. The learning rate is adapted using a \texttt{ReduceLROnPlateau} scheduler that monitors the validation loss and multiplies the learning rate by 0.5 whenever the validation loss fails to improve for 45 consecutive epochs. Early stopping with a patience of 50 epochs selects the final model.

For MMD, we use standard mini-batch training with batch size 512 and no learning-rate scheduler. Instead, we rely on early stopping with a patience of 10 epochs based on the validation loss. In all cases, the per-sample regression weights $w_i$ are the same and are described in Appendix~\ref{app:weights}. For the DA term, we fix the MMD loss weight to $\lambda = 0.5$ and the RBF kernel bandwidth to $\sigma = 1.0$. These values were selected on the basis of a small grid search over $\lambda \in \{0.1, 0.5, 0.9\}$ and $\sigma \in \{0.1, 1.0\}$: all combinations yielded similar regression losses, with more aggressive settings (larger $\lambda$ and smaller $\sigma$) slightly lowering the total training loss but leading to more volatile training and a stronger trade-off between domain alignment and regression accuracy. The choice $(\lambda, \sigma) = (0.5, 1.0)$ provides a stable, mid-range configuration that balances the contribution of the MMD term without pushing either hyperparameter to an extreme.

\begin{acknowledgements}
It's a pleasure to thank Taebong Jeong, Aswin Neelakandan for fruitful discussions. This work was supported by resources provided by the Pawsey Supercomputing Research Centre’s Setonix Supercomputer (https://doi.org/10.48569/18sb-8s43) and Acacia Object Storage (https://doi.org/10.48569/nfe9-a426), with funding from the Australian Government and the Government of Western Australia. The authors acknowledge the Texas Advanced Computing Center (TACC) at The University of Texas at Austin for providing computational resources that have contributed to the research results reported within this paper. 

This work made use of premier images captured by Subaru Telescope on the summit of Maunakea, Hawaii. We acknowledge the cultural, historical, and natural significance and reverence that Maunakea has for the indigenous Hawaiian community. We are deeply fortunate and grateful to share in the opportunity to explore the Universe from this mountain. The Hyper Suprime-Cam (HSC) collaboration includes the astronomical communities of Japan and Taiwan, and Princeton University. The HSC instrumentation and software were developed by the National Astronomical Observatory of Japan (NAOJ), the Kavli Institute for the Physics and Mathematics of the Universe (Kavli IPMU), the University of Tokyo, the High Energy Accelerator Research Organization (KEK), the Academia Sinica Institute for Astronomy and Astrophysics in Taiwan (ASIAA), and Princeton University. Funding was contributed by the FIRST program from Japanese Cabinet Office, the Ministry of Education, Culture, Sports, Science and Technology (MEXT), the Japan Society for the Promotion of Science (JSPS), Japan Science and Technology Agency (JST), the Toray Science Foundation, NAOJ, Kavli IPMU, KEK, ASIAA, and Princeton University.
KGL acknowledges support from JSPS Kakenhi grant Nos. JP18H05868, JP19K14755 and JP24H00241.
Kavli IPMU is supported by the World Premier International Research Center Initiative (WPI), MEXT, Japan. 
This work was performed in part at the Center for Data-Driven Discovery, Kavli IPMU (WPI).

\end{acknowledgements}

\bibliography{sample701}{}
\bibliographystyle{aasjournalv7}

\end{document}